\documentclass[12pt,a4paper]{article}
\usepackage[T1]{fontenc}
\usepackage[margin=2cm]{geometry}
\usepackage{amsmath}
\usepackage{accents, adjustbox, afterpage, amssymb, amsfonts, amsthm, array, booktabs, calc, caption, color, comment, datetime, dcolumn, enumitem, eurosym, float, geometry, graphics, graphicx, longtable, mathtools, mathbbol, multirow, pdflscape, pdfpages, ragged2e, rotating, sectsty, setspace, subcaption, subfiles, textpos, threeparttable, tikz, soul, xcolor, xparse
}
\usetikzlibrary{arrows.meta}
\usepackage[hang, flushmargin, bottom]{footmisc}

\usepackage[authoryear]{natbib}
\setcitestyle{authoryear,aysep={},open={(},close={)}}
\bibpunct{(}{)}{;}{a}{,}{,}

\usepackage[hypertexnames=false]{hyperref}
\hypersetup{
    pdffitwindow=false,colorlinks=true,linkcolor={red!50!black},citecolor={blue!50!black},urlcolor=black,linkbordercolor={white},
}
\usepackage{etoolbox}

\NewDocumentCommand{\hyref}{m O{}O{}}{\hyperref[#1]{#2 \ref{#1}#3}}
\setcitestyle{authoryear,aysep={},open={(},close={)}}
\bibpunct{(}{)}{;}{a}{,}{,}

\usepackage{etoolbox}

\makeatletter
\DeclareRobustCommand\citepos													
  {\begingroup\def\NAT@nmfmt##1{{\NAT@up##1's}}%
   \NAT@swafalse\let\NAT@ctype\z@\NAT@partrue
   \@ifstar{\NAT@fulltrue\NAT@citetp}{\NAT@fullfalse\NAT@citetp}}

\pretocmd{\NAT@citex}{%
  \let\NAT@hyper@\NAT@hyper@citex
  \def\NAT@postnote{#2}%
  \setcounter{NAT@total@cites}{0}%
  \setcounter{NAT@count@cites}{0}%
  \forcsvlist{\stepcounter{NAT@total@cites}\@gobble}{#3}}{}{}
\newcounter{NAT@total@cites}
\newcounter{NAT@count@cites}
\def\NAT@postnote{}

\def\NAT@hyper@citex#1{
  \stepcounter{NAT@count@cites}%
  \hyper@natlinkstart{\@citeb\@extra@b@citeb}#1%
  \ifnumequal{\value{NAT@count@cites}}{\value{NAT@total@cites}}
    {\if*\NAT@postnote*\else\NAT@cmt\NAT@postnote\global\def\NAT@postnote{}\fi}{}%
  \ifNAT@swa\else\if\relax\NAT@date\relax
  \else\NAT@@close\global\let\NAT@nm\@empty\fi\fi								
  \hyper@natlinkend}
\renewcommand\hyper@natlinkbreak[2]{#1}

\patchcmd{\NAT@citex}															
  {\ifNAT@swa\else\if*#2*\else\NAT@cmt#2\fi
   \if\relax\NAT@date\relax\else\NAT@@close\fi\fi}{}{}{}
\patchcmd{\NAT@citex}
  {\if\relax\NAT@date\relax\NAT@def@citea\else\NAT@def@citea@close\fi}
  {\if\relax\NAT@date\relax\NAT@def@citea\else\NAT@def@citea@space\fi}{}{}
\patchcmd{\NAT@cite}{\if*#3*}{\if*\NAT@postnote*}{}{}
\makeatother

\DeclareMathOperator*{\argmax}{arg\,max}

\DeclareMathOperator*{\supp}{supp}

\newcommand*\diff{\mathop{}\!\mathrm{d}}

\theoremstyle{plain}
\newtheorem{theorem}{Theorem}
\newtheorem{proposition}{Proposition}
\newtheorem{lemma}{Lemma}

\theoremstyle{definition}
\newtheorem{definition}{Definition}

\newtheorem{example}{Example}

\newtheoremstyle{newtheorem}{5pt}{2pt}{\itshape}{0pt}{\bfseries}{.}{.4em}{\thmname{#1}\thmnumber{ #2}\textnormal{\thmnote{ (#3)}}}

\newcommand\blfootnote[1]{\begingroup\renewcommand\thefootnote{}\footnote{#1}\addtocounter{footnote}{-1}\endgroup}

\usepackage[sf,sl,outermarks]{titlesec}
\newcommand{\addperiod}[1]{#1.}

\titleformat{\section}[block]
{\normalfont\Large\bfseries}{\thesection.}{.5em}{\Large\bfseries}
\titlespacing*{\section}{0pt}{*1.3}{*0.2}

\titleformat{\subsection}[block]
{\normalfont\large\bfseries}{\thesubsection.}{.5em}{\large\bfseries}
\titlespacing*{\subsection}{0pt}{*1}{*0}

\titleformat{\subsubsection}[block]
{\normalfont\large\bfseries}{\thesubsubsection.}{.5em}{\large\bfseries}
\titlespacing*{\subsubsection}{0pt}{*1}{*0}

\titleformat{\paragraph}[runin]
{\normalfont\bfseries}{}{0em}{\normalsize\bfseries\addperiod}
\titlespacing*{\paragraph}{0pt}{*.15}{*1}

\definecolor{mathematica1}{rgb}{0.368417, 0.506779, 0.709798}
\definecolor{mathematica2}{rgb}{0.880722, 0.611041, 0.142051}
\definecolor{mygreen}{RGB}{0,150,0}
\definecolor{myred}{RGB}{207,8,8}

\setlist[enumerate]{itemsep=0pt,topsep=2pt}

\newdateformat{DDMonthYYYY}{\THEDAY\, \monthname[\THEMONTH] \THEYEAR}

\usepackage{dsfont,fouriernc,charter,libertine}

\AtBeginDocument{
  \setlength\abovedisplayskip{4pt}
  \setlength\belowdisplayskip{4pt}
}

\defcitealias{FU2021105230}{Fu, Haghpanah, Hartline and Kleinberg (2021)}

\begin{document}
\thispagestyle{empty}
\setcounter{page}{0}
\setcounter{footnote}{0}
\renewcommand{\thefootnote}{\fnsymbol{footnote}}
~\vspace*{-2cm}\\
\begin{center}
    \Large\bfseries  
    Statistical Mechanism Design:\\[.5em] 
    Robust Pricing, Estimation and Inference
\end{center}

\makebox[\textwidth][c]{
    \begin{minipage}{1.2\linewidth}
        \Large\centering
        \begin{tabular}{clr}
            Duarte Gon\c{c}alves\footnotemark & 
            & 
            \qquad Bruno A. Furtado\footnotemark  \qquad \qquad 
        \end{tabular}
    \end{minipage}
}
\setcounter{footnote}{1}\footnotetext{
    \setstretch{1} Department of Economics, University College London; \hyperlink{mailto:duarte.goncalves@ucl.ac.uk}{\color{black}duarte.goncalves@ucl.ac.uk}.
    }
\setcounter{footnote}{2}\footnotetext{
    \setstretch{1} Department of Economics, Royal Holloway University of London; \hyperlink{mailto:bruno.dealbuquerquefurtado@rhul.ac.uk}{\color{black}bruno.dealbuquerquefurtado@rhul.ac.uk}.
    }
\blfootnote{
    We are grateful to Yeon-Koo Che, Tim Christensen, Mark Dean, Navin Kartik, Elliot Lipnowski, and Michael Woodford for their comments and suggestions.\\
	\emph{First posted draft}: 25 May 2020. \emph{This draft}: \DDMonthYYYY\today.
}
\vspace*{2em}

\setcounter{footnote}{0} \renewcommand{\thefootnote}{\arabic{footnote}}

\begin{center} \textbf{\large Abstract} \end{center}
\vspace*{1em}
\makebox[\textwidth][c]{
    \begin{minipage}{.85\textwidth}
        This paper tackles challenges in pricing and revenue projections due to consumer uncertainty.
        We propose a novel data-based approach for firms facing unknown consumer type distributions. 
        Unlike existing methods, we assume firms only observe a finite sample of consumers' types. 
        We introduce \emph{empirically optimal mechanisms}, a simple and intuitive class of sample-based mechanisms with strong finite-sample revenue guarantees.
        Furthermore, we leverage our results to develop a toolkit for statistical inference on profits. 
        Our approach allows to reliably estimate the profits associated for any particular mechanism, to construct confidence intervals, and to, more generally, conduct valid hypothesis testing. 
        \\\\
        \textbf{Keywords:} Robust mechanism design; Pricing; Hypothesis testing; Nonparametric estimation.\\
        \textbf{JEL Classifications:} D82; D42; C14; L12.
    \end{minipage}
}
\newpage

\section{Introduction}
\label{section:introduction}

Pricing and revenue projections are crucial elements of any firm's business plan. 
However, uncertainty about consumer willingness to pay makes both tasks challenging. 
While firms desire revenue guarantees, they also rely on future revenue projections for budgeting, inventory management, and capital investment decisions. 
These needs require reliable estimates of expected profits with confidence bounds.
This is especially true in the analogous setting of government procurement, where the procurement process may be subject to administrative rules and regulations, while at the same time having an influence on the overall government budget.
In such cases, having reliable projections of procurement results is crucial for planning and execution of public policy.

This paper proposes a novel data-based approach to address these challenges simultaneously. 
We consider a scenario where a firm faces uncertainty about the true distribution of consumer types, building on the classic framework of \citet{MaskinRiley1984RAND}. 
We allow the firm to design mechanisms (menus of price-quantity pairs) while acknowledging limitations in real-world information access.

Unlike many existing robust mechanism design approaches, we assume the firm observes only a finite sample drawn from the unknown distribution. 
This is a natural assumption insofar as it approximates the type of information that organizations tend to have at their disposal in practice, often as the result of market research \citep{LuskShogren2007Book}.
Accordingly, there has been growing interest in analyzing mechanism design with random samples, as exemplified by notable recent papers including \citetalias{FU2021105230}, \citet{Allouah2022}, and \citet{XieZhuShishkin2023WP}.

Our contribution to this literature is twofold. 
First, we provide a class of simple mechanisms with robust finite-sample revenue guarantees: \emph{empirically optimal mechanisms}. 
Second, we develop a toolkit to perform statistical inference on the profit obtained for any arbitrary mechanism, including the empirically optimal one. 
This enables a data-driven approach to the evaluation and comparison of different pricing strategies.

Empirically optimal mechanisms are remarkably simple to implement and boast notable revenue guarantees and robustness properties. 
Specifically, empirically optimal mechanisms simply involve consistently estimating the true distribution from a sample and implementing the mechanism that would be optimal for that estimated distribution. 
This gives rise to a data-based mechanism that is not only asymptotically optimal, but also strongly robust to small perturbations of the estimated distribution, owing to profit being Lipschitz continuous in the estimate of the distribution.
Furthermore, they induce finite-sample, exponential bounds for profit and regret (the difference to the optimal profit).

Equally important is the fact that that empirically optimal mechanisms serve as a tool for estimating and conducting inference on the profit obtained with any mechanism, including the optimal one. 
This allows the principal to obtain reliable confidence intervals for profits achievable with any mechanism, to compare and test profits and regret arising from different feasible pricing strategies, and also estimate the maximum achievable profit, which can be of interest to estimate the expected return to an investment or long-run profitability, as better information about the type distribution becomes available. 
In the analogous case of procurement from a firm, the ability to obtain reliable expenditure projections from any specific policy is particularly valuable, as the procurement process may be subject to idiosyncratic administrative constraints. 
Methodologically, we prove a novel envelope theorem to establish consistency and asymptotic normality of our proposed estimator via a functional delta method.
We then demonstrate the effectiveness of bootstrap implementations for conducting inference through Monte Carlo simulations, which suggest that the empirical coverage of our estimators approximates well the associated confidence intervals, even with relatively few samples.

Finally, we illustrate how our results on empirically optimal mechanisms extend to an auction setting. 
We consider the case where the firm auctions a single item to a finite number of risk-neutral bidders with independent private values drawn from the same distribution.
In particular, analogue versions of asymptotic optimality and profit and regret guarantees are shown to hold in this setting as well.

Empirically optimal mechanisms correspond to one of the simplest forms of statistically informed mechanism design: observing a sample, estimating a distribution, and implementing a mechanism that is optimal for the estimated distribution. 
As we demonstrate, this extremely simple approach boasts sound revenue guarantees and allows practitioners to estimate the maximum profit attainable. 
Our results on expected profit estimation enable confidence interval construction and hypothesis testing, valuable tools for practitioners and researchers. 
We believe this data-driven perspective on robust mechanism design holds promise for the broader study of mechanism design under model uncertainty.

\subsection{Related Literature}
\label{section:introduction:literature}

The most directly related literature studies robust mechanism design with a monopolist who does not form a prior about consumers' willingness to pay, but has access to a random sample from this distribution. 
Relatively recent papers with this set-up include \citet{ColeRoughgarden2014STOC14}, \citet{DHANGWATNOTAI2015318}, \citet{HuangMansourRoughgarden2018SIAMJComput}, \citet{GuoHuangZhang2020ACM}, \citet{FU2021105230}, \citet{Allouah2022}, and \citet{XieZhuShishkin2023WP}. 
Most of the work in this body of literature is concerned with either providing revenue guarantees based on a sample, or leveraging the sample to design a mechanism that maximizes profits conditional on the limited information available.

A sizeable literature has been dedicated to providing sample-based revenue guarantees in auctions. 
For example, \cite{HuangMansourRoughgarden2018SIAMJComput} give polynomial upper an lower bounds on the number of samples required to achieve a share $1 - \varepsilon$ of the optimal profit, assuming a single buyer with value drawn from a distribution with the monotone hazard rate property.
\cite{ColeRoughgarden2014STOC14} provide a similar characterization of the sample complexity of optimal mechanisms, but for multiple bidders. 
\cite{GuoHuangZhang2020ACM} improved on the results of the previous paper by providing upper and lower bounds that differ by at most a poly-logarithmic factor.

\cite{XieZhuShishkin2023WP} consider a different setting, where a monopolist has access to a random sample of data on buyers' valuations and covariates, and uses it to conduct third-degree price discrimination or to determine an uniform price.
They provide a lower bound on regret, measured as the minimax difference in expected revenue, for any sample-based strategy as a function of the sample size.
Moreover, they provide pricing strategies that attain this lower bound for both the third-degree price discrimination and uniform pricing regimes, and thus have the fastest possible rate of convergence to the optimal profit as the sample size increases.

Our first set of results is very close to this literature, as it provides revenue guarantees for the empirically optimal mechanism as a function of the sample size. 
Naturally, such results can also be inverted to obtain the number of samples that is needed to achieve a given share of the optimal profit with high probability. 
On the other hand, we significantly diverge from the papers cited above in the results concerning inference for any given mechanism. 
In a sense, the simplicity of the empirically optimal mechanism allows us to use it as an econometric tool to generate statistical analyses from the sample that are more detailed than simply providing bounds on regret.

Another strand of the literature is concerned with designing a mechanism that makes optimal use of the information contained in a sample of consumers' valuations. 
Early work by \cite{Segal2003AER} and \cite{BaligaVohra+2003} study a setting where samples are elicited from buyers' bids themselves, under different assumptions about the underlying distribution of valuations. 
They both propose mechanisms which induce consumers to reveal their own type, while simultaneously allowing the firm to use others' bids to infer the distribution and design the mechanism accordingly.
\cite{DHANGWATNOTAI2015318} follow \cite{BaligaVohra+2003} in designing a mechanism with a random reserve price drawn from the empirical distribution function of buyers' bids, and show that their Single Sample mechanism approximates that of an optimal auction.
\cite{FuHaghpanahHartlineKleinberg2020WP} posit that the possible distributions of bidders' types belongs to a finite set, and determine the minimum number of samples that an auctioneer needs to extract the full surplus without prior knowledge of the true distribution.

We are not concerned with making optimal use of the data to extract surplus. 
Rather, we study a particular mechanism and focus on providing a toolkit for conducting valid non-parametric inference in the mechanism design setting. 
Nevertheless, in \hyref{section:eom:rev}[Section] we quantify the extent to which the empirically optimal mechanism can be improved upon in terms of profit by any sample-based mechanism, and find that the scope for such improvements quickly vanishes as the sample size increases.

More generally, the present paper fits into the broader literature on mechanism design when the principal has perfect knowledge not about the whole distribution, as in the more standard models \citep[e.g.][]{MaskinRiley1984RAND}, but only about some features of this distribution.
With such information, the firm can then narrow down the set of possible distributions to consider and adopt a pricing strategy that maximizes the worst-case profit or minimizes the worst-case regret.
This approach tries to address the concern that the optimal mechanism is not robust to the firm having less than exact information 
on the distribution of consumers' willingness-to-pay, in line with the general research program of robust mechanism design.

In this broader literature, the papers which are closest to ours are \citeauthor{BergemannSchlag2008JEEA} \citeyearpar{BergemannSchlag2008JEEA, BergemannSchlag2011JET} and \citet{CarrascoLuzKosMessnerMonteiroMoreira2018JET}.
These papers model a firm that does not know the distribution of consumer types, but has access to imperfect information that allows it to refine the set of possible distributions.
Focusing on a linear specification, they assume the firm acts as if it faces an adversarial nature that chooses a distribution to maximize regret.\footnote{
    \citet{BergemannSchlag2011JET} also consider the case where nature minimizes profit and show the firm chooses a deterministic uniform pricing rule.
}
\citet{BergemannSchlag2008JEEA} assume that the firm knows only an upper bound for the support; \citet{BergemannSchlag2011JET} study the case where the firm also knows that the true distribution of consumers' willingness-to-pay is in a given neighbourhood of a given target distribution; \citet{CarrascoLuzKosMessnerMonteiroMoreira2018JET} posits that the firm knows either the first moment and an upper bound for the support of the distribution,\footnote{
    This is also the case in a closely related paper, \citet{CarrascoLuzMonteiroMoreira2018ET}, which relaxes the assumptions that consumers' utility function is linear in quantity and that the firm's cost is also linear.
} or the first two or three moments of the distribution.
These papers then characterize the regret-minimizing mechanisms, whereby the firm hedges against uncertainty by randomizing over prices. Our analysis shows that, in addition to its other desirable properties, the empirically optimal mechanism generates nearly minimal worst-case regret in the sense of these papers.

In another related paper, \citet{MadaraszPrat2017REStud} allow for a firm that is uncertain over both the distribution of types and the functional form of consumers' utility functions, while at the same time endowing the firm with a possibly misspecified benchmark model of consumer demand.
They provide a uniform bound on regret that depends on the distance between the firm's benchmark model and the truth, where the measure of distance between models is related to the largest absolute value of the difference of willingness-to-pay across all possible types and quantities.
Instead of looking at the worst-case scenario solution, the authors show that adjusting the pricing strategy that is optimal for the misspecified model in a specific manner involving this distance leads to a deterministic uniform bound on regret, similar to the probabilistic bound provided by the empirically optimal mechanism.

\vspace*{.5em}

A brief outline of the paper is as follows. 
\hyref{section:setup}[Section] introduces the main theoretical framework.
In \hyref{section:eom}[Section], we define our class of empirically optimal mechanisms and examine some of their main properties: asymptotic optimality and profit guarantees.
After exploring this particular class of mechanisms, in \hyref{section:inference}[Section] we turn to the question of providing a statistical toolkit to estimate and conduct inference on profit, including optimal expected profit.
\hyref{section:auctions}[Section] illustrates an extension of our results to the auction setting with independent private values.
Finally, we conclude with a discussion of specific suggestions for further work in \hyref{section:conclusion}[Section].
All omitted proofs are included in the \hyref{section:appendix:proofs}.

\section{Setup}
\label{section:setup}

Let $\Theta:=[\underline \theta, \overline \theta]\subset \mathbb R$, denote the set of feasible consumer types, which are distributed according to the cumulative distribution $F_0 \in \mathcal F$.
Denote by $\mathcal F$ the set of all distributions on $\Theta$ endowed with the supremum-norm ${\|\cdot\|}_\infty$, i.e., ${\|F\|}_\infty := \sup_{t \in \mathbb R} |F(t)|$ for all $F \in \mathcal F$.
Consumers' utility is given by $u(\theta,x,p)=v(\theta,x) - p$, where $\theta$ is the consumer's type, $x \in  X:=[0, \overline x]$ denotes quantity, and $p\in \mathbb R_+$ price.
We assume that $v$ is twice continuously differentiable, concave in $x$, supermodular, $v(\underline \theta,x)=v(\theta,0)=0$ for all $\theta$ and $x$, increasing in both arguments, and, wherever positive, strictly so.

The firm can choose a menu or mechanism $M$ from the set $\mathcal M$ of all compact menus $M' \subset X \times \mathbb R_+$ containing the element $(0,0)$.
These comprise pairs of quantity and prices that the consumers can choose, with the option of consuming nothing being always available.
We impose the further restriction that if $(x,0) \in M'$, then $x=0$; that is, the firm does not give away strictly positive quantities of the good for free.\footnote{
    This restriction facilitates \hyref{section:inference}[Section]'s inference exercise for an arbitrary fixed menu and is without loss of revenue from the firms' perspective.
}
The firm incurs a twice differentiable, convex, and strictly increasing cost for quantity sold, $c: X\to \mathbb R$, where $c(0)=0$.
When choosing menu $M\in \mathcal M$ and facing a distribution of consumer types $F \in \mathcal F$, the firm's expected profit $\pi(M,F)$ is given by
\begin{align*}
    \pi(M,F):=\int p(\theta)-c(x(\theta))\diff F(\theta),
\end{align*}
for some $(x(\theta),p(\theta)) \in \argmax_{(x,p) \in M} u(\theta,x,p)$, with ties broken in favor of the firm.

Our setup encompasses many of the variations in the literature, being mostly the same as that in the one buyer version of \citet{Myerson1981MOR} and \citet{MaskinRiley1984RAND}, except that there $c$ is assumed to be linear and $\mathcal F$ corresponds to the distributions with a strictly positive density. 
\citet{MussaRosen1978JET}, instead, assume that $v(\theta,x)=\theta\cdot x$ and specify $c$ to be strictly convex.
In \citet{BergemannSchlag2011JET} and \citet{CarrascoLuzKosMessnerMonteiroMoreira2018JET}, $v(\theta,x)=\theta\cdot x$ and $\mathcal F$ is a subset of distributions that satisfy some pre-specified conditions.\footnote{
    It does not include, for instance, the case where the firm's cost depends directly on the consumers' type as in Example 4.1 in \citet{Toikka2011JET}.
}

We consider the case where neither the firm nor the consumers know the true distribution of types, $F_0 \in \mathcal F$, which motivates the choice of dominant strategies as our solution concept. 
Instead, the firm has access to a sample $S^n=\left( \theta_i,\,i=1,...,n\right) \in \Theta^n$, $n \in \mathbb N$, where each $\theta_i$ is independently drawn from $F_0$. 
As discussed later (\hyref{section:conclusion}[Section]), such a sample can and will often arise from, for instance, market research previously conducted.
This gives rise to the problem of selecting a menu depending on the realized sample. 
Let $\mathcal S$ denote the set of all samples, $\mathcal S:=\bigcup_{n \in \mathbb N}\Theta^n$. 
A \emph{sample-based mechanism} is then a mapping $M_S: \mathcal S \to \mathcal M$, which selects a specific menu depending on the realized sample.

The robust mechanisms mentioned earlier can easily be adjusted to incorporate the information in the sample in order to estimate the features of the distribution that they assume to be known. 
For example, given a particular sample $S^n$, the firm can then estimate the support of the true distribution and implement the mechanism given in \citet{BergemannSchlag2008JEEA}. 
Alternatively, it can estimate the first few moments of the true distribution and implement the mechanism in \citet{CarrascoLuzKosMessnerMonteiroMoreira2018JET}. 
A natural question is then whether, given sampling uncertainty, these sample-based mechanisms would exhibit probabilistic robustness properties akin to the deterministic ones that hold when these features are perfectly known. 
This question is relevant in practice since \emph{any} information about an unknown distribution is usually obtained from finite data. 
We therefore take a more direct and, arguably, simpler approach that makes full use of the sample itself to ``learn'' about the underlying true distribution and inform mechanism choice.

\section{Empirically Optimal Mechanisms}
\label{section:eom}

In this section, we introduce the class of empirically optimal mechanisms.
This class of mechanisms is defined by two elements: an estimator of the true distribution and a mapping that takes each estimated distribution to a menu that would be optimal were the estimate to coincide with the true distribution.\footnote{
    Note that, if the seller did have a prior, $\mu \in \Delta (\Delta(\Theta))$, over the set of possible distributions, it would still be sufficient to consider only the expected distribution according to the seller's posterior, $\mathbb E_{\mu}[F|S^n] \in \mathcal F$, in order to determine which mechanism to choose. 
    That is, due to linearity of the profit function on the distribution,
    $\max_{M \in \mathcal M} \mathbb E_{\mu}[\pi(M,F)|S^n]=\max_{M \in \mathcal M} \pi(M,\mathbb E_{\mu}[F|S^n])$.
}

Let $\widehat{\mathcal F}$ denote the set of estimators that are consistent for $F_0$, that is, the set of estimators $\hat F$ such that (i) $\hat F: \mathcal S \to \mathcal F$; and (ii) ${\|\hat F(S^n)-F_0\|}_\infty \overset{p}{\to}0$, for any $F_0$ in $\mathcal F$.
Let $M^*:\mathcal F \to \mathcal M$ be a fixed selection from the set of optimal menus for every distribution, that is, $\forall F \in \mathcal F$, $M^*(F) \in \mathcal M^*(F):=\argmax_{M \in \mathcal M}\pi(M,F)$.
An \emph{empirically optimal mechanism} $\hat{M}^*$ is a sample-based mechanism that simply composes together a consistent estimator and a selection from the set of optimal menus, that is, $\hat{M}^*=M^* \circ\hat F$.
We refer to the set of empirically optimal mechanisms as ${\widehat{\mathcal M^*}}$.

While extremely simple, nothing ensures us that such a sample-based mechanism is either well-defined in general environments or that it constitutes a reasonable approach to pricing under uncertainty.
The purpose of this section is to address these issues and show that this simple approach to pricing delivers several desirable properties including strong probabilistic robustness guarantees.

\subsection{Existence}
\label{section:eom:existence}

In order to show that the set of empirically optimal mechanisms ${\widehat{\mathcal M^*}}$ is nonempty, we start by briefly noting that an optimal menu exists for any distribution $F \in \mathcal F$. 
Formally, we have that $\mathcal M^*(F)\ne \emptyset$ for all $F \in \mathcal F$ --- we include a formal proof of this statement in \hyref{appendix:lemma:existence:proof}. 
Since this implies that a selection $M^*: \mathcal F \to \mathcal M$ such that $M^*(F) \in \mathcal M^*(F)$ exists, and as there are consistent estimators for any $F_0\in \mathcal F$, the set of empirically optimal mechanisms ${\widehat{\mathcal M^*}}$ is nonempty. 
An example of a consistent (and unbiased) estimator for $F_0$ is the empirical cumulative distribution, defined as $\hat F(S^n)(\theta)=\frac{1}{n}\sum_{i=1}^n\mathbf 1_{\{\theta_i\leq \theta\}}$.
Moreover, for any $F_0 \in \mathcal F$, it has a uniform rate of convergence, that is, ${\|\hat F(S^n)-F_0\|}_\infty\overset{p}{\to}0$ (in fact, by the Glivenko--Cantelli theorem, uniform convergence occurs almost surely).

Note that under some specific conditions, an exact characterization of the set of optimal menus is known. 
For instance, if $v(\theta,x)=\theta\cdot x$ and $c(x)=\overline c \cdot x$, it is well-known that any optimal mechanism is $F$-almost everywhere equal to an indicator function $\mathbf 1_{\{p^*\geq \theta\}}$, where $p^*\in \argmax_{p \in \supp(F)}(p-\overline c)\cdot \int \mathbf 1_{\{p\geq \theta\}}\diff F(\theta)$. 
Such an explicit solution simplifies the problem of characterizing empirically optimal mechanisms dramatically. 
When, instead, $v$ is multiplicatively separable, and $F_0$ is absolutely continuous and has convex support, any optimal mechanism is almost everywhere equal to pointwise maximization of the ironed virtual value as shown in \citet{Toikka2011JET}. 
Hence, the problem of characterizing empirically optimal mechanisms can be made computationally tractable by ensuring not only consistency, but also absolute continuity and convex support of estimates $\hat F(S^n)$. 
Although the empirical cumulative distribution fails to be absolutely continuous, one such estimator $\hat F$ can be easily obtained by adopting any smooth interpolation of the empirical cumulative distribution, for example a linear interpolation, a cubic spline, or an interpolation relying on Bernstein polynomials.\footnote{
    In \hyref{appendix:other-proofs}, we provide a simple proof for the fact that linear interpolations retain the uniform convergence (and therefore consistency) properties of the empirical distribution.
    See \citet{BabuCantyChaubey2002JSPI} and \citet{Leblanc2011AnnInstStatMath} for details on interpolation of the empirical cumulative distribution using Bernstein polynomials.
}

\subsection{Asymptotic Optimality}
\label{section:eom:asymptotic-optimality}

Having defined our class of empirically optimal mechanisms, we establish in this section that they are asymptotically optimal: the realized expected profit given the mechanism converges in probability to the optimal expected profit as the sample size grows.

Such convergence is not guaranteed for arbitrary sample-based mechanisms, even for those that have desirable robustness properties, as it requires that the sample-based mechanism makes full use of the sample. 
For instance, if the mechanism relies only on statistics such as estimates for a finite number of moments or the support of the distribution, then it is immediate that it will not, in general, converge in probability to the optimal expected profit. 
Relying on an estimator for the true distribution itself (an infinite-dimensional parameter) is then key to obtaining asymptotic optimality.

We start by making an important observation:
\begin{lemma}
    \label{lemma:profit:lipschitz}
    For any $M \in \mathcal M$, $\pi(M,F)$ is Lipschitz continuous in $F \in \mathcal F$, with a Lipschitz constant $L$ that does not depend on $M$.
\end{lemma}

We defer the proof to \hyref{appendix:lemma:profit:lipschitz}, but highlight the main steps here. 
The proof of \hyref{lemma:profit:lipschitz}[Lemma] first makes use of the revelation principle to focus on the elements of any arbitrary menu that are payoff relevant, as these are given by a bounded non-decreasing function, and hence of bounded variation. 
Then, we appeal to a result that provides an upper bound on Riemann--Stieltjes integrals of functions of bounded variation to obtain the result. 
In particular, we find a Lipschitz constant of at most 
\begin{align*}
    L=2\left(v(\overline \theta, \overline x)+(\overline \theta - \underline \theta) \cdot \max_{\theta' \in \Theta}v_1(\theta', \overline x)+c(\overline x)\right).
\end{align*}

For every $F \in \mathcal F$, define the firm's value function as \[\Pi(F):=\sup_{M \in \mathcal M}\pi(M,F).\]
\hyref{lemma:profit:lipschitz}[Lemma] leads to a further result, this time regarding (Lipschitz) continuity of the value function:
\begin{lemma}
    \label{lemma:optimal-profit:lipschitz}
    $\Pi$ is Lipschitz continuous, with Lipschitz constant $L$.
\end{lemma}
\begin{proof}
    For any $F, G \in \mathcal F$, 
    $\displaystyle
        |\Pi(F)-\Pi(G)|=\left|\sup_{M \in \mathcal M}\pi(M,F)-\sup_{M \in \mathcal M}\pi(M,G)\right|\leq \sup_{M \in \mathcal M}|\pi(M,F)-\pi(M,G)|\leq L\cdot {\|F-G\|}_\infty.
    $
\end{proof}

Finally, the desired result of asymptotic optimality of our class of empirically optimal mechanisms follows immediately:
\begin{proposition}
    \label{proposition:asymptotic-optimality}
    Let $\hat{M}^*$ be an empirically optimal mechanism given by $\hat{M}^*=M^* \circ \hat F$.
    Then, $|\pi(\hat{M}^*(S^n),F_0)-\Pi(F_0)|\overset{p}{\to}0$.
\end{proposition}
\begin{proof}
    By \hyref{lemma:profit:lipschitz}[Lemmas] and \ref{lemma:optimal-profit:lipschitz}, we have that 
        $|\pi(\hat{M}^*(S^n),F_0)-\Pi(F_0)|\leq |\pi(\hat{M}^*(S^n),F_0)-\pi(\hat{M}^*(S^n),\hat F(S^n))| + |\pi(\hat{M}^*(S^n),\hat F(S^n))-\Pi(F_0)|
        \leq L\cdot {\|\hat F(S^n)-F_0\|}_\infty+|\Pi(\hat F(S^n))-\Pi(F_0)|\leq 2L\cdot {\|\hat F(S^n)-F_0\|}_\infty\overset{p}{\to}0$.
\end{proof}

\hyref{proposition:asymptotic-optimality}[Proposition] provides a simple justification for using an empirically optimal mechanism to guide the firm's pricing strategy: as the sample size grows large, such sample-based mechanisms deliver an expected profit close to the optimal one. 
We stress the minimal informational assumptions made. 
In particular, this result does not depend on the firm knowing the support of the true distribution $F_0$ ex ante, as the empirically optimal mechanism is defined making use only of the estimated cumulative distribution and there are consistent estimators that require no assumptions on (and in fact, asymptotically learn) the support of $F_0$. 
Moreover, as $\Pi(F_0)-\pi(\hat{M}^*(S),F_0)\leq 2 L {\|\hat F(S)-F_0\|}_\infty$, we conclude that empirically optimal mechanisms are robust in a sense akin to \citet{BergemannSchlag2011JET}, since for any $\varepsilon>0$, samples inducing ${\|\hat F(S)-F_0\|}_\infty<\varepsilon$ imply that $\Pi(F_0)-\pi(\hat{M}^*(S^n),F_0)\leq 2 L \varepsilon$.

\subsection{Robustness Properties}
\label{section:eom:rev}

While some sample-based implementations of existing robust mechanisms would not be asymptotically optimal, it is possible that others would.
Thus, an obvious question is: Do empirically optimal mechanisms provide robustness guarantees with finite samples that render them especially appealing?
In this section, we argue that this is indeed the case.

Robustness properties of mechanisms regard worst-case scenarios.
The existing literature has focused on two main properties.
One corresponds to the worst-case profit that the firm can expect given that the true distribution lies in a specific set $A \subseteq \mathcal F$.
This is in part motivated by appealing to the characterization of preferences exhibiting ambiguity aversion by \citet{GilboaSchmeidler1989JMathEcon}, which entails a maxmin representation, whereby the decision-maker (here, the firm) evaluates each act (mechanism) by assuming the worst-case payoff.
The robustness of a specific mechanism according to this criterion is then given by the lower bound on expected profit it can attain, $\min_{F \in A}\pi(M,F)$.
The second robustness criterion that has been considered in the literature depends on the notion of regret: How much profit the firm may be forgoing by committing to mechanism $M$ when the true distribution is $F_0$, that is, $R(M,F_0):=\Pi(F_0)-\pi(M,F_0)$.

A simple implication of \hyref{lemma:profit:lipschitz}[Lemmas] and \ref{lemma:optimal-profit:lipschitz} is that we can immediately obtain probabilistic bounds on regret and on how far the realized expected profit may be from the profit the firm expects to obtain given its estimated distribution.
\begin{proposition}
    \label{proposition:regretbounds}
    Let $\hat{M}^*=M^*\circ \hat F$ be an empirically optimal mechanism.
    Suppose that $\hat F \in \widehat{\mathcal F}$ is such that $\forall S^n \in \mathcal S$, $\mathbb P({\|\hat F(S^n)-F_0\|}_\infty >\delta)\leq p(n,\delta)$ for some function $p:\mathbb N \times \mathbb R_+ \to [0,1]$.
    Then,
        (i) $\mathbb P\left(|\pi(\hat{M}^*(S^n),\hat F(S^n))-\pi(\hat{M}^*(S^n),F_0)|>\delta\right)\leq p(n,\delta/L)$; and
        (ii) $\mathbb P\left(R(\hat{M}^*(S^n),F_0)>2\delta\right)\leq p(n,\delta/L)$,
    where $L$ denotes the Lipschitz constant from \hyref{lemma:profit:lipschitz}[Lemma].
\end{proposition}

While \hyref{proposition:regretbounds}[Proposition] is a trivial observation, it enables the firm to obtain strong, non-asymptotic probabilistic bounds on both profit and regret whenever basing an empirically optimal mechanism on an estimator $\hat F$ with specific properties. 
Whenever the firm knows an upper bound $\overline \theta$ for the support of the true distribution, $L$ can be obtained in a way that depends exclusively on known constants and these probabilistic profit and regret guarantees can be computed explicitly.
The next two examples illustrate how this result can be applied by focusing on estimators $\hat F$ with well-known properties, such as the empirical cumulative distribution and smooth interpolations of it.

\begin{example}
    \label{example:ecdf}
    Let $\hat{M}^*=M^*\circ \hat F$ be any empirically optimal mechanism such that $\hat F$ denotes the cumulative distribution estimator.
    By the Dvoretzky--Kiefer--Wolfowitz inequality \citep{DvoretzkyKieferWolfowitz1956AnnMathStat} with \citeauthor{Massart1990AnnProb}'s \citeyearpar{Massart1990AnnProb} constant, we have that 
    \begin{align*}
        \mathbb P({\|\hat F_n-F_0\|}_\infty>\delta)&\leq 2\exp\left(-2n\delta^2\right)
    \end{align*}
    and hence $p(n,\delta)=2\exp\left(-2n\delta^2\right)$.
    Then, \hyref{proposition:regretbounds}[Proposition] applies and we can obtain regret lower than $2\delta$ with probability of at least $1-p(n,\delta)$ and a confidence bound with range $2 \delta$ such that the true expected profit differs from $\pi(\hat{M}^*(S^n),\hat F(S^n))$ by less than $\delta$ also with probability greater than $1-p(n,\delta)$.
    \qed
\end{example}

\begin{example}
    \label{example:iecdf}
    Suppose that $\mathcal F$ is restricted to the set of absolutely continuous distributions on $\Theta$, of which $F_0$ is known to be an element of, and that $v$ is multiplicatively separable in $\theta \in \Theta$ and $x\in X$.
    If we were to constrain $\hat F(S)$ to also be absolutely continuous, admitting a strictly positive density and having convex support, an analytic characterization of $M^*(\hat F(S))$ is known, given by pointwise maximization of the ironed virtual value given $\hat F(S)$.\footnote{
        \citet{Toikka2011JET} has shown that for any absolutely continuous distribution $F$ on $\Theta$ with a strictly positive density, the set of maximizers $\mathcal M^*(F)$ pointwise maximize $\bar{J}(\theta)v(\theta,x) - c(x)$, where $\bar{J}(\theta)$ is the ironed version of $J(\theta) := \theta - \frac{1 - F(\theta)}{f(\theta)}$.
    }
    This would then simplify the computational cost of finding the optimal mechanism.
    Take $\hat F$ to be any interpolation of the empirical cumulative distribution that results in a valid distribution function that is absolutely continuous and has convex support, such as the linear interpolation (see \hyref{lemma:iecdf}[Lemma] in the appendix).
    Note that given that $F_0$ is atomless by assumption, $\mathbb P\left(\exists k,\ell \in \{0,1,...,n-1\}\; : \, s_{k}=s_{\ell}\right)=0$ $\forall n \in \mathbb N$, and thus the linear interpolation is well-defined with probability 1. 
    Furthermore, for any such interpolation $\hat F$, with probability 1, the estimate given by $\hat F(S^n)$ differs from the empirical cumulative distribution $\hat F^{E}(S^n)$ by at most $1/n$ at any given point.
    Hence,
    \begin{align*}
        &\mathbb P({\|\hat F(S^n)-F_0\|}_\infty>\delta)\leq \mathbb P({\|\hat F^{E}(S^n)-F_0\|}_\infty+{\|\hat F(S^n)-\hat F^{E}(S^n)\|}_\infty>\delta)\\
        &\leq \mathbb P({\|\hat F^{E}(S^n)-F_0\|}_\infty>\delta-1/n)\leq 2\exp\left(-2n(\delta-1/n)^2\right),
    \end{align*}
    where, again, the last inequality follows from the Dvoretzky--Kiefer--Wolfowitz inequality with \citeauthor{Massart1990AnnProb}'s constant. 
    It follows that $p(n,\delta)=2\exp\left(-2n(\delta-1/n)^2\right)$.
    As $\hat F \in \widehat {\mathcal F}$, \hyref{proposition:regretbounds}[Proposition] applies and the regret and confidence bounds are obtained.
    \qed
\end{example}

As this next example shows, under some assumptions on the true distribution $F_0$ one can even obtain not only non-asymptotic, but also deterministic (i.e., non-probabilistic) regret and confidence bounds.

\begin{example}
    \label{example:kernel}
    Suppose that it is known that the true distribution $F_0$ admits a density $f_0$ with total variation bounded by $B<\infty$ and has support contained in $[0,1]$.\footnote{This can be generalized to some closed interval in $\mathbb R$.}
    Let $V_B$ denote the set of all densities on $[0,1]$ with total variation bounded by $B$.
    Take any kernel density estimator, given by
    \begin{align*}
        \hat f(S^n)(u) = \frac{1}{n h_n}\sum_{i=1}^n K\left(\frac{u-\theta_i}{h}\right),
    \end{align*}
    where $h$ is a smoothing parameter such that $h \to 0$ as $n\to \infty$, $\theta_i$ denotes the value of observation $i$, $K\geq 0$ and $\int K(u)\diff u=1$.
    From theorem 3 in \citet{Datta1992AnnStat}, we know that, for any such kernel density estimator $\hat f_n$, one has
    \begin{align*}
        \sup_{f \in V_B}\|\hat f(S^n)-f\|_1 \leq (2 B+1)k_1 h + {\left(\frac{k_2}{n h}\right)}^{1/2},
    \end{align*}
    where $k_1:=\int |u| K(u)\diff u$ and $k_2:=\int K^2(u)\diff u$.
    Define $q(n,h):=(2 B+1)k_1 h + {\left(\frac{k_2}{n h}\right)}^{1/2}$ and let $\hat F(S^n)(\theta):=\int \mathbf 1_{u\leq \theta}\hat f(S^n)(u)\diff u$ denote the estimated cumulative distribution.\\
    Note that there are several kernel density estimators available such that 
    \[k_1 =\\ \int |u| K(u) \diff u < \infty\,\,\text{ and }\,\,k_2=\int K^2(u)\diff u<\infty,\] 
    e.g. if $K$ is the uniform kernel, triangle, Epanechnikov, among others.\\
    In order for $\hat F$ to belong to $\widehat{\mathcal F}$ we need for it to (1) be uniformly consistent and (2) to have compact support.
    There are several ways to achieve this, namely by relying on a kernel such that $\int_{|u|>\delta}K(u)\diff u=0$ for some finite $\delta>0$ --- which is satisfied by most of the standard kernels, namely the ones cited above.
    For any such kernel, $k_1, k_2<\infty$ and $\forall S^n \in \mathcal S$, $\text{supp } \hat F(S^n) \subseteq \Theta$ is compact, by choosing lower and upper bounds $\underline \theta,\; \overline \theta$ appropriately.\\
    Therefore,
    \begin{align*}
        {\|\hat F(S^n)-F_0\|}_\infty&=\sup_{\theta \in \Theta}\left|\hat F(S^n)(\theta)-F_0(\theta)\right|=  \sup_{\theta\in \Theta}\left|\int \mathbf 1_{u\leq \theta}\left(\hat f(S^n)(u)-f_0(u)\diff s\right)\right|\\
        &\leq \sup_{\theta\in \Theta}\int \mathbf 1_{u\leq \theta}|\hat f(S^n)(u)-f_0(u)| \diff s \leq  {\|\hat f(S^n)-f_0\|}_1 \leq q(n,h),
    \end{align*}
    where ${\|\cdot\|}_1$ denotes the $L^1$ norm, i.e. ${\|f\|}_1:=\int |f(u)|\diff u$.
    It follows immediately that if $h\to 0$ and $n \cdot h \to \infty$, ${\|\hat F(S^n)-F_0\|}_\infty\to 0$ and, consequently, $\hat F \in \widehat{\mathcal F}$. The argument above implies that
        $R(\hat{M}^*(S^n),F_0)\leq q(n,h)/2 L$ 
        and $|\pi(\hat{M}^*(S^n),\hat F(S^n))-\pi(\hat{M}^*(S^n),F_0)|\leq q(n,h)/L$.
    As such, if $f_0$ has total variation bounded by $B$, we can obtain non-probabilistic bounds on regret and expected profit when implementing an empirically optimal mechanism based on kernel density estimation.
    \qed
\end{example}

Other sample-based mechanisms could potentially yield even stronger robustness properties.
However, for any sample-based mechanism $M_S$, one has
\begin{align*}
    R(M_S(S^n),F_0)-R(\hat{M}^*(S^n),F_0)&=\pi(M_S(S^n), F_0)-\pi(\hat{M}^*(S^n), F_0)\\
    &\leq \Pi(F_0)-\pi(\hat{M}^*(S^n), F_0)
    =R(\hat{M}^*(S^n),F_0).
\end{align*}
As the examples above show, choosing $\hat F$ appropriately ensures that $\mathbb P(R(\hat{M}^*(S^n),F_0)>\delta)$ declines exponentially with $n$.
Hence, the gains in profit and regret guarantees from implementing alternative pricing policies are modest at best, in a formal sense.

To conclude this section, we note that \hyref{proposition:regretbounds}[Proposition] can instead be used to determine how many samples the firm requires in order to obtain specific robustness guarantees when relying on empirically optimal mechanisms.
That is, another reading of \hyref{proposition:regretbounds}[Proposition] is that the firm only needs at most $N$ samples --- where $N$ is the smallest integer such that $\alpha\geq p(N,\delta/L)$ --- to secure at most $2\delta$ of regret with probability $1-\alpha$. Alternatively, the same $N$ samples provide a (conservative) confidence interval for profit with range $2\delta$ and confidence level of $1-\alpha$.
This results in a non-asymptotic sample complexity bound for a specific class of sample-based mechanisms.
In contrast to the sample-complexity bounds obtained in \citet{HuangMansourRoughgarden2018SIAMJComput}, which pertain to the share of the optimal profit that the firm is able to secure, $R(M,F_0)/\Pi(F_0)$, we focus on bounding regret directly and our bounds are not asymptotic, that is, they hold for finite samples.

\section{Inference and Robustness}
\label{section:inference}

While the revenue and regret guarantees derived in the previous section are useful, it is often crucial for a firm to to be able to obtain consistent projections for the profit for any given pricing strategy.
This consistency is essential for informing various business activities, such as budget planning and inventory management, based on reliable profit estimations.
In this section, we show how to obtain consistent and unbiased estimates and conduct inference on the expected profit.

Our results enable unbiased and consistent estimation of and inference on the expected profit not only of empirically optimal mechanisms but of any given mechanism $M \in \mathcal M$. 
Moreover, we show how empirically optimal mechanisms serve a special purpose, in that they can be used to estimate and conduct inference on the optimal expected profit. 
With these tools, one can provide confidence intervals for the expected profit with specific asymptotic coverage for any one mechanism, calculate probabilistic bounds for regret, or test whether one mechanism yields a higher expected profit than another.

\subsection{Expected Profit}
\label{section:inference:rob}
An immediate consequence of Lipschitz continuity of the firm's profit function with respect to the distribution is that, for any mechanism and any consistent estimator of the distribution of types, one can consistently estimate the expected profit that such a mechanism would generate.

\begin{proposition}
    \label{proposition:profit:consistency}
    For any true distribution $F_0 \in \mathcal F$, consistent estimator $\hat F \in \widehat{\mathcal F}$, and mechanism $M \in \mathcal M$, we have $\pi(M,\hat F(S^n)) \overset{p}{\to} \pi(M,F_0)$. 
    Moreover, if $\hat F$ is an unbiased estimator such as the empirical distribution function, $\mathbb E\left[\pi(M,\hat F(S^n))\right]=\pi(M,F_0)$, that is, the plug-in estimator is also unbiased.
\end{proposition}
\begin{proof}
    By \hyref{lemma:profit:lipschitz}[Lemma], we have that, $\forall \hat F \in \widehat{\mathcal F}$ and $\forall M \in \mathcal M$,
    $\displaystyle
        |\pi(M,\hat F(S^n))-\pi(M,F_0)|\leq L\|\hat F(S^n)-F_0\|_\infty \overset{p}{\to} 0.
    $
    Moreover, if $\hat F$ is an unbiased estimator of $F_0$, by linearity of $\pi(M,F)$ in $F$, we have that $\mathbb E\left[\pi(M,\hat F(S^n))\right]=\pi(M,\mathbb E\left[\hat F(S^n)\right]) = \pi(M,F_0)$.
\end{proof}

An important aspect of estimation is the ability to conduct inference.
For instance, the firm could be interested in using statistical inference in order to compare different mechanisms, that is, to test whether a specific mechanism would deliver a higher expected profit than another.
Another possible application would be to obtain valid confidence intervals for the expected profit under a particular mechanism, as it is an arguably crucial tool for the development of routine business activities such as drawing up budgets under different scenarios, with varying degrees of confidence.

While the firm could potentially derive confidence intervals for the expected profit by adjusting \hyref{proposition:regretbounds}[Proposition] and \hyref{example:ecdf}[Example] to the mechanism it is considering, these bounds would exhibit two drawbacks when used for this purpose.
First, they require knowledge of $\overline \theta$, the upper bound on the type distribution.
Second, and more critically, they generally do not provide the correct asymptotic coverage, that is, they would be exceedingly conservative.

In order to address these drawbacks, we suggest a simple estimation procedure that does yield asymptotically valid inference.
First, we focus on the empirical distribution function as our estimator, since it admits a functional central limit theorem (by Donsker's theorem) when properly centered and rescaled.
Then, because profit is linear and continuous in the type distribution, it is Fr\'{e}chet differentiable, with derivative given by $\dot \pi_M(\cdot) = \pi(M,\cdot)$.\footnote{
    The Fr\'{e}chet derivative $\dot \pi_M$ is defined on the space of functions on $\Theta$ of bounded variation endowed with the supremum-norm.
    We refer to the appendix for details.
}
One can thus obtain the asymptotic distribution of our consistent estimator for the expected profit $\pi(M,\hat F(S^n))$ by appealing to a simple functional Delta method result.

\begin{theorem}
    \label{theorem:profit:asymptotic-normality}
    Let $\hat F$ denote the empirical distribution estimator.
    Then, $\forall M \in \mathcal M$ and $\forall F_0 \in \mathcal F$, we have that 
    $
    \displaystyle \sqrt{n}\left(\pi(M,\hat F(S^n)) - \pi(M,F_0)\right) \overset{d}{\to} N(0,\sigma_{M,F_0}^2)
    $,
    where $\sigma_{M,F_0}^2 :=\mathbb E\left[{\left(\dot\pi_{M}(\delta_{\theta}-F_0)\right)}^2\right] = \mathbb E\left[\left(\pi(M,\delta_{\theta}-F_0)\right)^2\right]$, $\theta \sim F_0$, and $\delta_{\theta}$ the cumulative distribution associated with a Dirac measure at $\theta$.
\end{theorem}

\hyref{theorem:profit:asymptotic-normality}[Theorem] states that the distribution of the empirical process, 
\[ G_n:=\sqrt n\left(\pi(M,\hat F(S^n))-\pi(M,F_0)\right),\] converges weakly to $N(0,\sigma_{M,F_0}^2)$; proof is given in \hyref{appendix:theorem:profit:asymptotic-normality}.
A question then arises of how to estimate, in practice, the asymptotic distribution in a consistent manner, as it depends on the unknown distribution $F_0$. 
We provide two alternatives. 
One option is the use of a plug-in estimator for $\sigma_{M,F_0}^2$. 
This can be done directly --- as the functional dependence of $\sigma_{M,F_0}^2$ on $F_0$ is known and a consistent estimate for $F_0$ is readily available --- or by following other plug-in methods as those in \citet{Shao1993AnnStat}. 
Another option is to rely on the classical bootstrap to approximate the distribution of $G_n$ by the distribution of $\hat G_n:=\sqrt n\left(\pi(M,\hat F(S^n_B))-\pi(M,\hat F(S^n))\right)$, conditional on $S^n$, where $S^n_B$ denotes the resampling of $n$ observations from $S^n$ with uniform weights. 
That this approach does in fact consistently estimate the limiting distribution was shown by \citet[Theorem 4]{Parr1985StatProbL}.

We note that other bootstrap methods would also yield consistent estimates, such as subsampling (bootstrap without replacement) \citep{PolitisRomano1994AnnStat} or Jackknife procedures \citep{Parr1985JRRS}.
Moreover, given that the Fr\'{e}chet derivative of profit is sufficiently well-behaved, under some smoothness assumptions on the true distribution $F_0$, smoothed versions of the bootstrap \citep{CuevasRomo1997AnnInstStatMath} can also be considered.

\subsection{Optimal Profit and Regret}
\label{section:inference:optimal}

We now extend our statistical inference results,  highlighting how empirically optimal mechanisms can be used to provide a consistent and asymptotically normal estimator for optimal profit.

Given that the firm's value function is Lipschitz continuous in the distribution and that empirically optimal mechanisms attain the optimal profit for a consistent estimate of the true distribution, it is easy to see that one can then use them as a tool to consistently estimate the optimal profit that the firm would obtain, were it to know $F_0$.
We formalize this observation as follows:

\begin{proposition}
    \label{proposition:optimal-profit:consistency}
    For any true distribution $F_0 \in \mathcal F$ and empirically optimal mechanism $\hat{M}^*$ given by $\hat{M}^*=M^* \circ \hat F$, $\pi(\hat M^*(S^n),\; \hat F(S^n)) \overset{p}{\to} \Pi(F_0)$.
\end{proposition}
\begin{proof}
    Similarly to \hyref{proposition:profit:consistency}[Proposition], we again have that, by \hyref{lemma:optimal-profit:lipschitz}[Lemma], $\Pi$ is Lipschitz continuous and therefore, 
    $\displaystyle
        |\pi(\hat M^*(S^n),\hat F(S^n))-\Pi(F_0)|=|\Pi(\hat F(S^n))-\Pi(F_0)|\leq L \|\hat F(S^n)-F_0\|_\infty \overset{p}{\to} 0.
    $
\end{proof}

It is less straightforward that one could take an approach to conducting inference on the optimal profit similar to that derived for fixed mechanisms.
Specifically, this would require proving that the firm's value function $\Pi$ is also Fr\'{e}chet differentiable.\footnote{
    In fact, the now standard functional Delta method requires only the weaker notion of Hadamard differentiability; see, e.g., \citet[ch. 3.9]{VaartWellner1996}. 
    However, the stronger notion of Fr\'{e}chet differentiability has the benefit of allowing us to bypass the measurability complications that arise when using weaker notions.
}
We confirm that indeed such an approach is valid by proving an interesting technical result in this generalized Maskin--Riley setup: an envelope theorem for the firm's value function.
In other words, our next result --- for which the proof is given in \hyref{appendix:theorem:optimal-profit:frechet} --- shows that the value function is Fr\'{e}chet differentiable at any distribution $F \in \mathcal F$ and that its Fr\'{e}chet derivative coincides with that of the expected profit at $F$ with the optimal menu for $F$.

\begin{theorem}[Envelope Theorem]
    \label{theorem:optimal-profit:frechet}
    $\Pi$ is Fr\'{e}chet differentiable at all $F \in \mathcal F$.
    Moreover, its Fr\'{e}chet derivative at $F$ is given by $\dot \Pi_F=\dot \pi_{M_F}$, $\forall M_F \in \mathcal M^*(F)$.
\end{theorem}

The proof can be found in \hyref{appendix:theorem:optimal-profit:frechet}. 
We note that \hyref{theorem:optimal-profit:frechet}[Theorem] constitutes a much stronger result than the standard envelope theorem.\footnote{
    The results in \citet{MilgromSegal2002Ecta} yield only directional derivatives and are insufficient to deliver asymptotic normality of our estimator.
}
It is exactly owing to delivering on a stronger notion of differentiability that we obtain asymptotically valid inference.
Specifically, defining the empirical process $\hat G_n:=\sqrt n\left(\Pi(\hat F(S^n_B))-\Pi(\hat F(S^n))\right)$, conditional on $S^n$, an adapted version of \hyref{theorem:profit:asymptotic-normality}[Theorem] ensues:

\begin{theorem}
    \label{theorem:optimal-profit:asymptotic-normality}
    Let $\hat F$ denote the empirical distribution estimator. 
    Then, $\forall F_0 \in \mathcal F$, 
    $\displaystyle
        \sqrt{n}\left(\Pi(\hat F(S^n)) - \Pi(F_0)\right) \overset{d}{\to} N(0,\sigma_{F_0}^2),
    $ 
    where $\sigma_{F_0}^2=\mathbb E\left[{\left(\dot\Pi_{F_0}(\delta_{\theta}-F_0)\right)}^2\right] = \mathbb E\left[{\left(\pi(M_F,\delta_{\theta}-F_0)\right)}^2\right]$ and $\theta \sim F_0$. 
    Moreover, $\hat G_n \overset{d}{\to} N(0,\sigma_{F_0}^2)$.
\end{theorem}

In this case, opposite to the case of inference under a fixed mechanism, we do not have a valid (consistent) plug-in estimator for $\sigma_{F_0}^2$, which depends on $F_0$ and, more problematically, also on an optimal mechanism under the distribution $F_0$, $M_0 \in \mathcal M^*(F_0)$.
An important argument in favour of a bootstrap approach to estimating the asymptotic distribution in this case is that it bypasses this issue.

Under some conditions, it may make sense to rely on different estimators.
For instance, as discussed in \hyref{example:iecdf}[Example], when $v(\theta,x)$ is multiplicatively separable and $F_0$ is known to be absolutely continuous and with compact and convex support, the functional form of the solution is exactly known. 
This allows for a drastic simplification of the problem from a computational point of view since it dispenses with the hurdle of finding the optimal mechanism for a given distribution. 
Especially in the context of implementing a bootstrap approach, the gains can be substantial. 
However, an estimate of the ironed virtual value depends on a suitable estimate of the density $f$. 
Therefore, we find it especially relevant that, when $F_0$ is known to be absolutely continuous, one can use as an estimator the simple linear interpolation of the empirical distribution discussed earlier to obtain a bootstrap estimator for the asymptotic distribution. 
Further, we note that this extremely simple approach is not only consistent for the true distribution $F_0$, but also for its density.

\begin{proposition}
    \label{proposition:optimal-profit:liecdf}
    Let $\hat F$ denote the linear interpolation of the empirical distribution estimator.
    Then, for any absolutely continuous $F_0 \in \mathcal F$, 
        (1) $\sqrt{n}{\left(\Pi(\hat F(S^n)) - \Pi(F_0)\right)}\overset{d}{\to} N(0,\sigma_{F_0}^2)$, where $\sigma_{F_0}^2=\mathbb E\left[{\left(\dot\Pi_{F_0}(\delta_{\theta}-F_0)\right)}^2\right]$;
        (2) $\hat G_n \overset{d}{\to} N(0,\sigma_{F_0}^2)$; and
        (3) ${\|\hat f(S^n) - f_0\|}_1 \xrightarrow{p} 0$, where $\hat f(S^n)$ and $f_0$ denote the Radon-Nikodym derivatives of $\hat F(S^n)$ and $F_0$, respectively.
\end{proposition}

Estimating the optimal expected profit may be relevant for investment decisions, as it provides an upper bound on the return of a given investment.
Furthermore, these results can also be used to estimate regret. 
As regret, $R(M,F)$, is given by $R(M,F)=\Pi(F)-\pi(M,F)$, it is Fr\'{e}chet differentiable at any distribution $F\in \mathcal F$, for any fixed $M \in \mathcal M$, as the sum of Fr\'{e}chet differentiable functionals is itself Fr\'{e}chet differentiable. 
Then, using similar arguments as those in \hyref{proposition:optimal-profit:consistency}[Proposition] and \hyref{theorem:optimal-profit:asymptotic-normality}[Theorem], one can conduct inference on regret. 
Consequently, for any mechanism $M \in \mathcal M$, one can not only obtain asymptotically valid probabilistic bounds for expected profit, but also for regret.

\subsection{Simulation Evidence}
\label{section:inference:simulations}

To conclude this section, we present empirical evidence on the finite sample properties of our estimators.

\afterpage{
\begin{table}[tp]
    \caption{Empirical Coverage Frequencies}\label{table:profit-a}
    \centering \small
    \begin{subtable}{\textwidth}
        \centering
        \caption{Profit with Fixed Mechanism}
        \begin{tabular}{cr|lll|lll|lll}
              &       & \multicolumn{3}{c|}{Beta(1/4,1/4)} & \multicolumn{3}{c|}{Unif(0,1)} & \multicolumn{3}{c}{Beta(4,4)} \\ \hline
              & $1-\alpha$ & .90   & .95  & .99  & .90   & .95  & .99  & .90   & .95  & .99 \\ \hline
            \multirow{3}[0]{*}{$N$} 
        	  & 500   & .891 &  .944 &  .983 &  .892 &  .946 &  .987 &  .895 &  .943 &  .988 \\
              & 1,000 & .901 &  .934 &  .985 &  .901 &  .946 &  .986 &  .896 &  .953 &  .985 \\
              & 2,500 & .909 &  .954 &  .989 &  .903 &  .953 &  .989 &  .889 &  .948 &  .985 \\
        \end{tabular}%
    \end{subtable}
    \vspace*{.5em}
    \begin{subtable}{\textwidth}
        \centering
        \caption{Optimal Profit}
        \begin{tabular}{cr|lll|lll|lll}
              &       & \multicolumn{3}{c|}{Beta(1/4,1/4)} & \multicolumn{3}{c|}{Unif(0,1)} & \multicolumn{3}{c}{Beta(4,4)} \\ \hline
              & $1-\alpha$ & .90   & .95  & .99  & .90   & .95  & .99  & .90   & .95  & .99 \\ \hline
            \multirow{3}[0]{*}{$N$} 
        	  & 500   & .881 &  .933 &  .989 &  .894 &  .945 &  .983 &  .895 &  .950 &  .987 \\
              & 1,000 & .886 &  .945 &  .985 &  .888 &  .943 &  .981 &  .894 &  .941 &  .984 \\
              & 2,500 & .910 &  .952 &  .988 &  .890 &  .941 &  .988 &  .884 &  .933 &  .984 \\
        \end{tabular}
    \end{subtable}
    \vspace*{.5em}
    \begin{minipage}{\textwidth}
        \footnotesize
        Note: This table shows the frequency with which the estimated confidence interval with asymptotic coverage of $1-\alpha$ contained the true expected profit, $\pi(M,F_0)$ in the case of a fixed mechanism and $\Pi(F_0)$ in the case of the optimal expected profit.
        The fixed mechanism corresponds to uniform pricing at $1/2$.
        The estimated confidence interval followed a centred bootstrap procedure with 1,000 samples redrawn with replacement from the original sample with 1,000 iterations, for each sample size $N$.
    \end{minipage}
\end{table}
}

We conduct Monte Carlo simulations on the empirical coverage of the confidence intervals for expected profit under a fixed mechanism --- uniform pricing, with the price set at $1/2$ --- and for the optimal expected profit, using empirically optimal mechanisms for the empirical distribution.
We use the approximation obtained by classic bootstrapping ($N$ out of $N$), which we have shown to be asymptotically valid.
We show simulation results for confidence levels $\alpha \in \left\{.1,.05,.01\right\}$, with varying sample size $N$.
For each sample size, we draw $1,000$ samples and, for each sample, we estimate the confidence interval by drawing $1,000$ bootstrap samples from the original sample.

\afterpage{
\begin{figure}[tp]
    \centering
    \begin{subfigure}{\textwidth}
        \begin{subfigure}{.5\textwidth}
            \includegraphics[width=\textwidth]{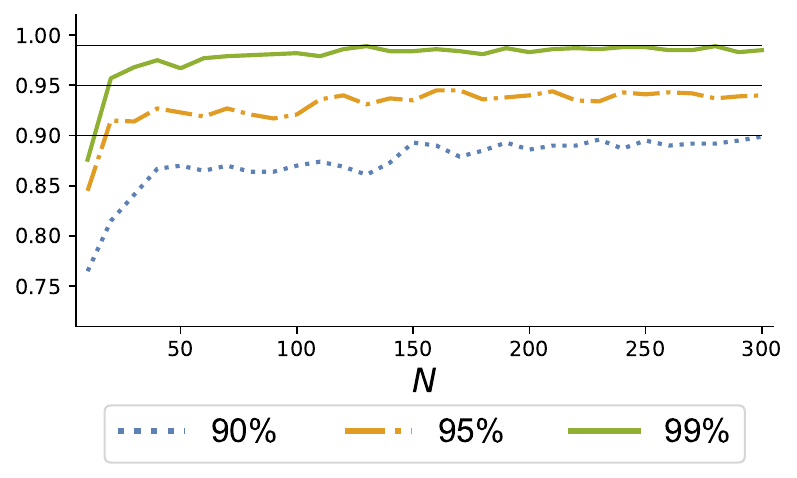}
            \renewcommand\thesubfigure{\alph{subfigure}1}
            \caption{Profit with Fixed Mechanism}\label{figure:F0-pix-a}
        \end{subfigure}
        \begin{subfigure}{.5\textwidth}
            \includegraphics[width=\textwidth]{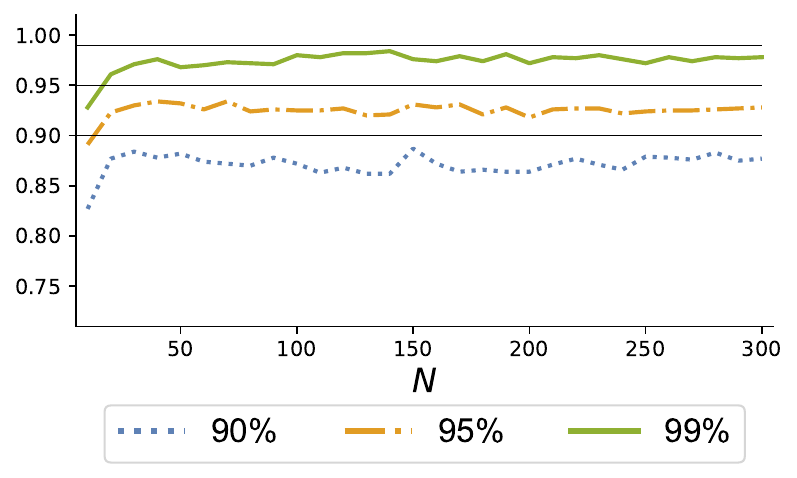}
            \addtocounter{subfigure}{-1}
            \renewcommand\thesubfigure{\alph{subfigure}2}
            \caption{Optimal Profit}
            \label{figure:F0-pio-a}
        \end{subfigure}
        \addtocounter{subfigure}{-1}
        \caption{Beta(1/4,1/4)}
        \label{figure:F0-a}
    \end{subfigure}
    \vspace*{.5em}
    \centering
    \begin{subfigure}{\textwidth}
        \begin{subfigure}{.5\textwidth}
            \includegraphics[width=\textwidth]{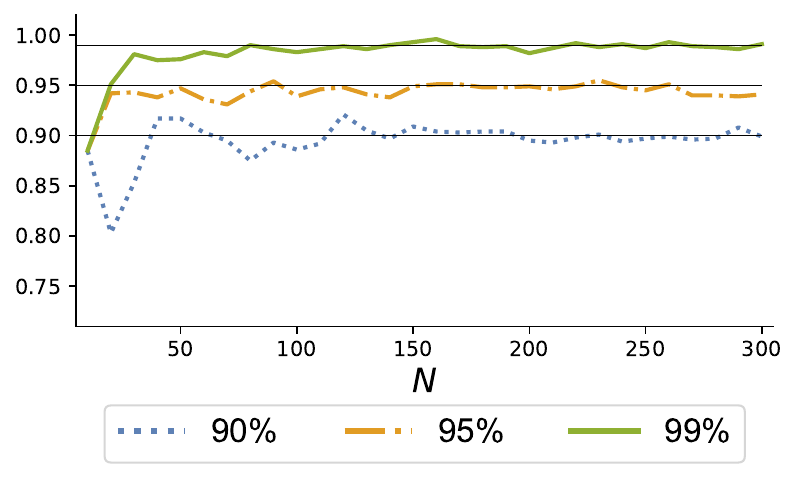}
            \renewcommand\thesubfigure{\alph{subfigure}1}
            \caption{Profit with Fixed Mechanism}\label{figure:F1-pix-a}
        \end{subfigure}
        \begin{subfigure}{.5\textwidth}
            \includegraphics[width=\textwidth]{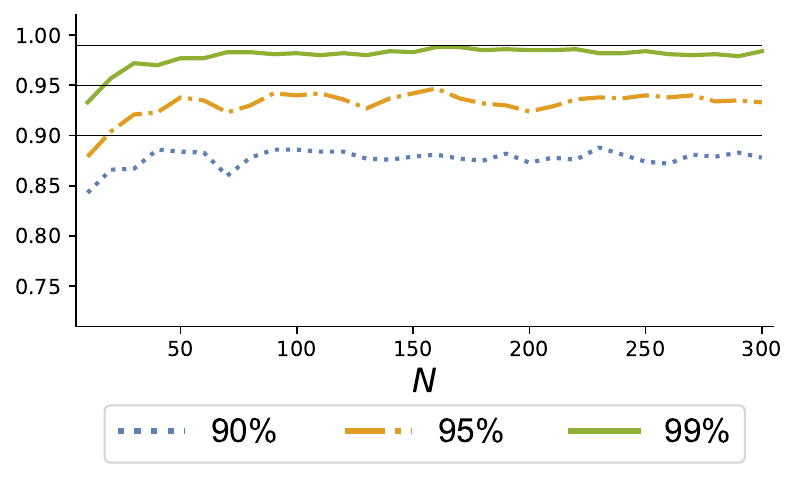}
            \addtocounter{subfigure}{-1}
            \renewcommand\thesubfigure{\alph{subfigure}2}
            \caption{Optimal Profit}
            \label{figure:F1-pio-a}
        \end{subfigure}
        \addtocounter{subfigure}{-1}
        \caption{Unif(0,1)}
        \label{figure:F1-a}
    \end{subfigure}
    \vspace*{.5em}
    \centering
    \begin{subfigure}{\textwidth}
        \begin{subfigure}{.5\textwidth}
            \includegraphics[width=\textwidth]{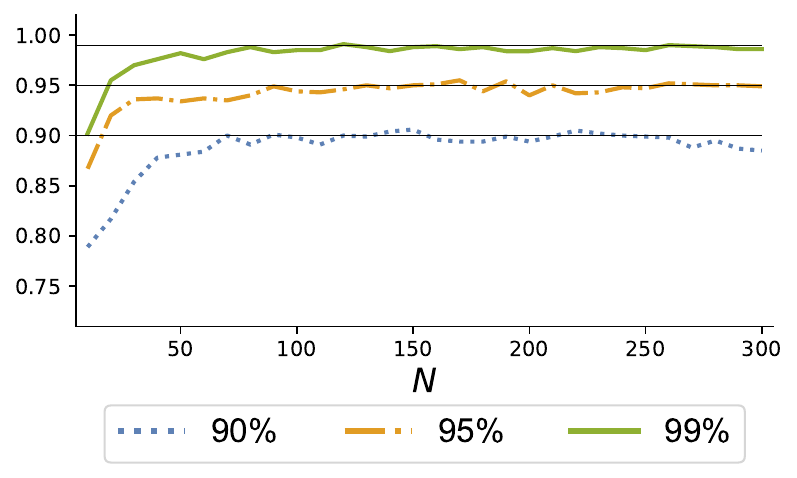}
            \renewcommand\thesubfigure{\alph{subfigure}1}
            \caption{Profit with Fixed Mechanism}
            \label{figure:F2-pix-a}
        \end{subfigure}
        \begin{subfigure}{.5\textwidth}
            \includegraphics[width=\textwidth]{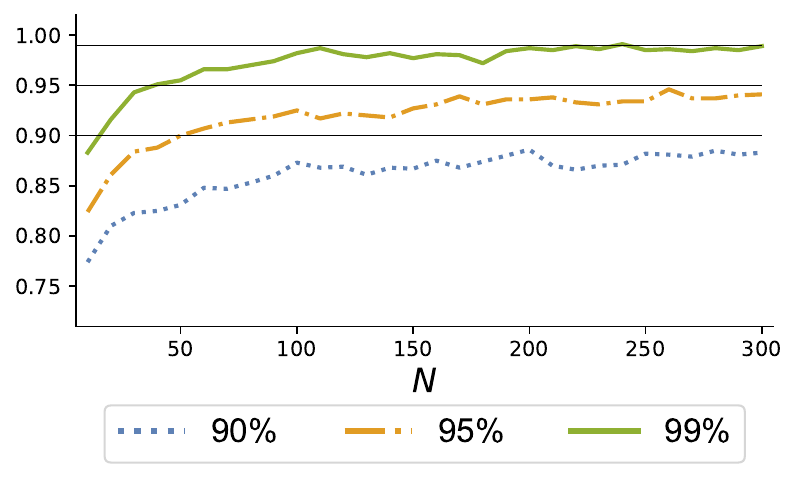}
            \addtocounter{subfigure}{-1}\renewcommand\thesubfigure{\alph{subfigure}2}
            \caption{Optimal Profit}
            \label{figure:F2-pio-a}
        \end{subfigure}
        \addtocounter{subfigure}{-1}
        \caption{Beta(4,4)}
        \label{figure:F2-a}
    \end{subfigure}
    \vspace*{.25em}
    \caption{Empirical Coverage Frequencies}
    \label{figure:profit-a}
    \vspace*{.25em}
    \begin{minipage}{\textwidth}
        \footnotesize\setstretch{1}
        Note: This figure shows the frequency with which the estimated confidence interval with asymptotic coverage of $1-\alpha =.9,\,.95,\,.99$ contained the true expected profit, $\pi(M,F_0)$ in the case of a fixed mechanism and $\Pi(F_0)$ in the case of the optimal expected profit.
        The procedure is as described in the note to \hyref{table:profit-a}[Table].
        Sample size $N$ varies between 10 and 300 with increments of 10 observations.
        The fixed mechanism corresponds to uniform pricing at $1/2$.
    \end{minipage}
\end{figure}
}

We focus on the case where consumers have quasilinear-linear utility and the unit cost is normalized to zero, as in \citet{BergemannSchlag2011JET} and \citet{CarrascoLuzKosMessnerMonteiroMoreira2018JET}.
We show results for three different parameterizations of $F_0$ relying on the Beta distribution: Beta(1/4,1/4), Uniform(0,1) and Beta(4,4).\footnote{
    The empirical coverage results were consistent across other parameterizations and using Beta mixtures or mixtures with degenerate distributions.
}

\afterpage{
\begin{figure}[t]
    \centering
    \includegraphics[width=.65\textwidth]{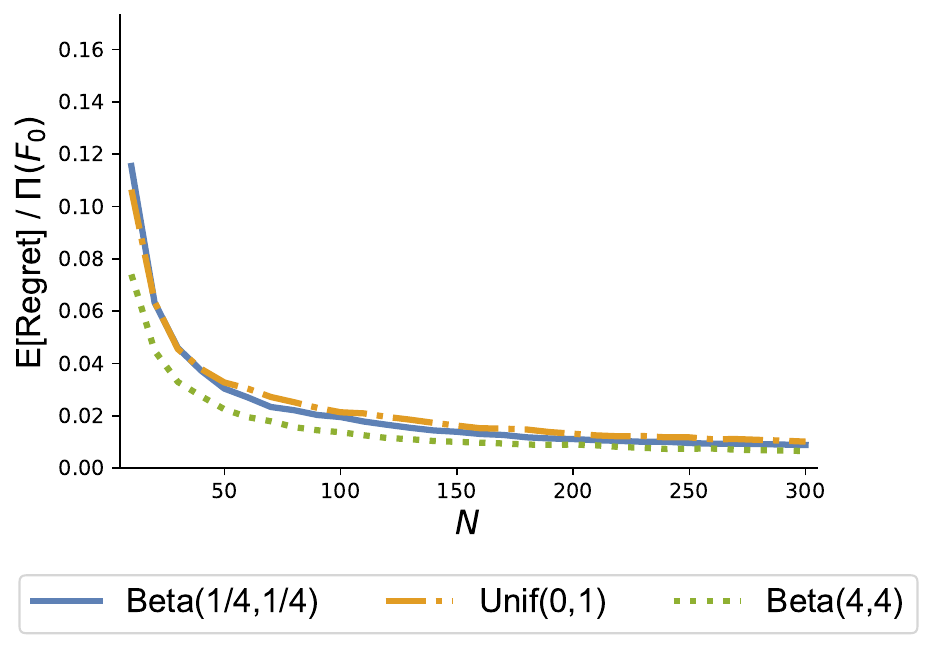}
    \caption{Regret of the Empirically Optimal Mechanism as a share of Optimal Profit}
    \label{figure:regret}
    \begin{minipage}{\textwidth}
        \footnotesize\setstretch{1}
        Note: This figure shows the average regret of the empirically optimal mechanism as a fraction of the optimal expected profit under the true distribution $F_0$.
        The average is taken over 1,000 samples for each sample size $N$ between 10 and 300 with increments of 10 observations.
    \end{minipage}
\end{figure}
}

In \hyref{table:profit-a}[Table] we present evidence for the empirical coverage frequency at sample sizes of 500, 1,000 and 2,500.
As is immediate upon inspection of the table, our estimators have extremely good finite sample properties, with the empirical coverage frequencies being very close to the theoretical asymptotic coverage probability, regardless of which of the three distributions is considered.
We also investigated the behaviour of our estimators under small samples.
As \hyref{figure:profit-a}[Figure] shows, they fare reasonably well for sample sizes between 50 and 300.

We also considered the regret incurred by adopting an empirically optimal mechanism that depends on the empirical distribution with finite samples.
As illustrated in \hyref{figure:regret}[Figure], we empirically study the average regret as a share of the optimal expected profit, that is, $\left(\Pi(F_0)-\pi(\hat M^*(S^n), F_0)\right)/\Pi(F_0)$.
The average is taken across 1,000 samples of varying size in increments of 10 observations.
Even with just 50 observations, the empirically optimal mechanism on average attains regret that is under 4\% of the optimal expected profit.
For the purpose of comparison, the robust mechanism in \citet[Section 5.1]{CarrascoLuzKosMessnerMonteiroMoreira2018JET}, relying on an estimate of the mean and assuming knowledge of the upper bound of the distribution, exhibits average regret no lower than 20\% of the optimal expected profit under any of the three distributions we consider.\footnote{
    We observe that for the minimax regret distribution derived in \citet{CarrascoLuzKosMessnerMonteiroMoreira2018JET}, by construction, the empirically optimal mechanism will attain the optimal profit with probability one and regardless of the number of samples, and, therefore, also attain the minimal regret.
}

\section{Extension to Single-Item Auctions}
\label{section:auctions}

Before concluding, we discuss how to apply some of the insights developed in this paper to the related setting of single-unit auctions.
In particular, we show how simple empirically optimal mechanisms in this context exhibit some of the desirable robustness features shown in \hyref{section:eom}[Section].

Suppose that the firm has a single item to auction to $M\geq 2$ bidders.
The firm values the item at $c>0$, and each bidder $i=1,...,M$ is risk-neutral and values the item at $\theta_i$, drawn independently from the distribution $F_0$.
To make matters simple, we assume that $F_0$ is absolutely continuous with convex and compact support.
In such case, it is well-known that revenue equivalence holds and that a second-price auction with a reserve price is optimal for the firm, with bidders disclosing their types.
Then, the optimal reserve price when the type distribution $F$ satisfies the same assumptions solves
    $\displaystyle\max_{r \in \Theta}\pi(r,F)$,
where $\pi(r,F)=\int_r^{\overline \theta} \diff F_{(2;M)}$ and $F_{(2;M)}$ denotes the distribution of the second-highest willingness-to-pay, given a distribution of types $F$ and $M$ bidders. That is, 
$$F_{(2;M)}(\theta)=M\cdot F(\theta)^{M-1}(1-F(\theta)) + F(\theta)^{M}.$$

Consider the case where the firm has access to a sample of $n$ observations drawn from $F_0$ and the reserve price is set before bids are submitted.\footnote{
    The same arguments apply when the reserve price is secret and takes into account the bids submitted, as these would just translate into a larger sample of $n+M$ observations.
}
Similar to before, denote an empirically optimal reserve price $\hat r^*$ as the composition of a consistent estimator, $\hat F$, of the true distribution ${F_{0}}$, based on the realized sample $S^n$, and a selection from the set of reserve prices that are optimal for $F$, $r^*$.

The next proposition provides an analogue of \hyref{proposition:asymptotic-optimality}[Propositions] and \ref{proposition:regretbounds} to this specific setting:
\begin{proposition}\label{proposition:auctions}
    Let $\hat{r}^*$ be an empirically optimal reserve price given by $\hat{r}^*=r^* \circ \hat F$.
    Then, $|\pi(\hat{r}^*(S^n),F_0)-\Pi(F_0)|\overset{p}{\to}0$.
    Moreover, if $\hat F \in \widehat{\mathcal F}$ is such that $\forall S^n \in \mathcal S$, $\mathbb P({\|\hat F(S^n)-F_0\|}_\infty >\delta)\leq p(n,\delta)$ for some function $p:\mathbb N \times \mathbb R_+ \to [0,1]$, then,
        (i) $\mathbb P\left(|\pi(\hat{r}^*(S^n),\hat F(S^n))-\pi(\hat{r}^*(S^n),F_0)|>\delta\right)\leq p(n,\delta/L)$; and
        (ii) $\mathbb P\left(R(\hat{r}^*(S^n),F_0)>2\delta\right)\leq p(n,\delta/L)$,
    where $L=2M(M-1)$ and $\Pi(F):=\sup_{r \in \Theta}\pi(r,F)$.
\end{proposition}

The key insight is that the expected profit is linear in distribution of the second-order statistic and that this in turn is Lipschitz continuous in the distribution of types, $\|F_{(2;M)}-G_{(2;M)}\|_\infty\leq 2M(M-1)\|F-G\|_\infty$.
For $\hat r^*$ to be empirically optimal, though, we must have that $\hat F(S^n)$ is absolutely continuous and has convex and compact support.
Similarly to \hyref{example:iecdf}[Example], when $\hat F$ is the linearly interpolated empirical distribution we have that $p(n,\delta)=2\exp(-2 n(\delta-1/n)^2)$, delivering regret and confidence bounds.\footnote{
    \citet{ColeRoughgarden2014STOC14} provide alternative sample-complexity bounds for this problem, characterizing the asymptotic number of samples needed to achieve $(1-\epsilon)$ share of the optimal profit.
}

As this application illustrates, our results on the robustness properties of empirically optimal mechanisms extend naturally to auction settings. 
There is, however, a natural limitation in extending our results on inference: expected profit is linear in the distribution of the second-order statistic, not in the distribution of types themselves.

\section{Concluding Remarks}
\label{section:conclusion}

This paper has studied two separate but related questions. 
The first is how a firm should price when uncertain about the distribution of consumers' willingness-to-pay. 
The second is how to conduct inference regarding the expected profit, both under any fixed pricing strategy and for the optimal profit.
When the firm has access to a sample of consumers' valuations, we have shown that adopting an extremely simple approach --- estimating the distribution using the sample and then pricing optimally for the estimated distribution --- yields attractive robustness properties, in particular obtaining probabilistic lower bounds both for the expected profit and for regret.
On the other hand, we provided a toolkit to conduct inference for the profit.
This enables practitioners to obtain confidence intervals not only for expected profit but also, for example, for the difference in profit that two different mechanisms induce.
More generally, this allows for a data-based approach to robust mechanism design, where robustness properties are inferred from available data.

One important concern that we have not discussed is how to obtain such a sample.
In order to elicit types (i.e., generate a sample), the firm could for instance conduct market research and elicit such a sample by means of a mechanism that induces each type to self-select to a different pair of quantity and price.\footnote{
    We are assuming consumers are myopic or that they do not gain from misrepresenting their type.
    Incentive compatibility regarding dynamic incentives if the firm can preclude surveyed customers from purchasing the item in the future, e.g., by supplying the good, in the case of unit demand. 
} 
Indeed, this is common practice: beyond gathering qualitative information about consumers' preferences from surveys, focus groups, or interviews, many consultancy firms also experimentally elicit consumers' types using incentive-compatible mechanisms in what is known in the industry under the umbrella term of `experimental auctions' --- even though sometimes it simply corresponds to the BDM mechanism \citep{BeckerDegrootMarschak1964BehaSc}.\footnote{
    See \citet{LuskShogren2007Book} for a comprehensive review of their use in marketing and management. 
    In the industry, Kantar, Ipsos, and Nielsen are examples of leading consultancy firms providing these services.
} 
This market research can then lead to the original sample or expand an existing sample --- in which case a cost-benefit analysis on the net value of acquiring additional observations may come into play.

One interesting avenue for future research considers the dynamical choice of information combining sampling and experimentation.
While, under some conditions, optimal experimentation also asymptotically attains the optimal profit \citep[see, e.g.,][]{AghionBoltonHarrisJullien1991REStud}, the implied cost is often distinct.
Naturally, the interplay between these two sources of information will dictate when and how much market research a firm does to optimise its pricing, compared to learning-by-doing.

\section{References}
\vspace*{-0em}\setlength{\bibhang}{0pt}
\bibliographystyle{aea}
{\setstretch{1.15}\setlength{\bibsep}{.0em plus .0ex}
\bibliography{smd.bib}}

\renewcommand{\thesection}{Appendix}
\renewcommand{\thesubsection}{Appendix \Alph{subsection}}
\renewcommand{\thesubsubsection}{\thesubsection.\arabic{subsubsection}}

\section{~}
\label{section:appendix:proofs}

\subsection{Existence of optimal mechanisms for arbitrary type distributions}
\label{appendix:lemma:existence:proof}

\begin{lemma}
    \label{lemma:existence}
    $\mathcal M^*(F)\ne \emptyset$ for all $F \in \mathcal F$.
\end{lemma}
\begin{proof}
    Define $\overline p := \max_{(\theta, x) \in \Theta \times X} v(\theta, x)$ and endow the power set of $X \times [0,\overline p]$ with the Hausdorff metric. 
    It is well-known that with the topology induced by the Hausdorff metric, the set of all compact subsets of $X \times [0,\overline p]$, denoted by $\mathcal K(X \times [0, \overline p])$, is itself compact. 
    Define the set of mechanisms $\mathcal M_{IR}$ as
    \[\mathcal M_{IR} := \left\{M \in \mathcal M : \max_{(x,p) \in M} p \leq \overline p\right\} \subseteq \mathcal K(X \times [0,\overline p]).\]
    
    Clearly, $\mathcal M_{IR}$ is closed, and therefore also compact. 
    Since $v(\cdot,\cdot)$ is continuous and clearly so is the mapping $\Theta \times \mathcal M_{IR} \ni (\theta, M) \mapsto M \cup (0,0)$, we have by the Berge's maximum theorem that the correspondence $\mathcal M_{IC}: \Theta \times \mathcal M_{IR}\rightrightarrows X \times [0,\overline p]$ such that 
    \begin{align*}
        (\theta, M) &\mapsto \mathcal M_{IC}(\theta,M)=\argmax_{(x,p)\in M\cup(0,0)} v(\theta,x) - p
    \end{align*}
    is non-empty, compact valued and upper hemicontinuous. 
    Applying the maximum theorem once again, we obtain that the value function $\gamma: \Theta \times \mathcal M_{IR} \to \mathbb R$ where 
    \begin{align*}
        (\theta, M) &\mapsto \gamma(\theta,M) = \max_{(x,p) \in \mathcal M_{IC}(\theta,M)} p - c(x)
    \end{align*}
    is upper semicontinuous, i.e., for any $(\theta_0, M_0) \in \Theta \times \mathcal M_{IR}$, $\limsup_{(\theta, M) \to (\theta_0,M_0)} \gamma(\theta, M) \leq \gamma(\theta_0, M_0)$. 
    Now, since $\gamma$ is bounded, we can apply (reverse) Fatou's Lemma to obtain that for any $M_0 \in \mathcal M_{IR}$,
    \[\limsup_{M_n \to M_0} \int \gamma(\theta,M_n) \diff F(\theta) \leq \int \limsup_{M_n \to M_0} \gamma(\theta, M_n) \diff F(\theta) \leq \int \gamma(\theta,M_0) \diff F(\theta).\]
    
    This proves that the mapping $(F,M) \mapsto \int \gamma(\cdot, M) \diff F$ is upper semicontinuous in $M$ for every $F \in \mathcal F$. 
    Therefore, since $\mathcal M_{IR}$ is compact, the extreme value theorem guarantees the existence of $M^* \in \argmax_{M \in \mathcal M_{IR}} \pi(M,F) = \argmax_{M \in \mathcal M} \pi(M,F)$ for all $F \in \mathcal F$, where the latter equality comes from the fact that $(x,p) \in X \times (\overline p, \infty)$ will never be chosen over $(0,0)$.
\end{proof}

\subsection{Proof of \hyref{lemma:profit:lipschitz}[Lemma]}
\label{appendix:lemma:profit:lipschitz}

We first state two results that will prove useful in determining the Lipschitz continuity of $\pi(M,F)$ in $F\in \mathcal F$. 
Let $\mathcal P:=\left\{P={\theta_0,\ldots,\theta_{n_P}}:\, \underline \theta=\theta_0\leq \theta_1\leq \cdots \leq \theta_{n_P}=\overline \theta\right\}$ and $V(f):=\sup_{P \in \mathcal P}\sum_{i=0}^{n_P-1}|f(\theta_{i+1})-f(\theta_i)|$, where 
$V(f)$ denotes the total variation of a function $f:\Theta\to \mathbb R$.
We have: 
\begin{lemma}[Beesack-Darst-Pollard Inequality \citep{DarstPollard1970ProcAMS,Beesack1975TheRockyMountainJournalofMathematics}]
    \label{lemma:boundsintegral}
    Let $f,g,h$ be real-valued functions on a compact interval $I=[a,b] \subset \mathbb R$, where $h$ is of bounded variation with total variation $V(h)$ on $I$ and such that $\int_a^b f \diff g$ and $\int_a^b h \diff g$ both exist.
    Then, 
    \begin{align*}
        m \int_{a}^{b}f \diff g + V(h) \sup_{a\leq a'\leq b'\leq b} \int_{a'}^{b'}f \diff g &\geq \int_{a}^{b}hf\diff g  
        \geq m \int_{a}^{b}f \diff g + V(h) \inf_{a\leq a'\leq b'\leq b} \int_{a'}^{b'}f \diff g,
    \end{align*}
    where $m= \inf\left\{h(x): x \in I\right\}$.
\end{lemma}

The following is a standard result in the mechanism design literature based on \citet{Mirrlees1971REStud} and \citet{MilgromSegal2002Ecta}.

\begin{lemma}
    \label{lemma:revelation}
    The choices from a menu $M \in \mathcal M$ satisfy $\displaystyle(x(\theta),p(\theta)) \in \argmax_{(x,p) \in M\cup\{(0,0)\}}u(\theta,x,p)$ if, and only if,  $x(\cdot)$ is nondecreasing, $x(\underline \theta)=0$ and $p(\theta)=v(\theta,x(\theta))-\int_{\underline \theta}^{\theta}v_1(s,x(s)) \diff s$.
\end{lemma}

For any $F \in \mathcal F$, we can thus rewrite the problem --- with some abuse of notation --- by choosing directly a function $x$ from the set $\mathcal X$, where 
    $\displaystyle\mathcal X:=\left\{x: [\underline \theta,\overline \theta] \to X,\, x\text{ is nondecreasing and } x(\underline \theta)=0\right\},$
in order to maximize profit, given by
\begin{align*}
    \pi(x,F):=\int_\Theta \left(v(\theta,x(\theta))-\int_{\underline \theta}^{\theta}v_1(s,x(s))\diff s-c(x(\theta))\right)\diff F(\theta).
\end{align*}

Consider the normed vector space $(BV(\Theta), {\|\cdot\|}_\infty)$, where $BV(\Theta) := \{g: \Theta \to \mathbb R \;|\; V(g) < \infty\}$. 
For any fixed $M \in \mathcal M$, consider its corresponding allocation function $x \in \mathcal X$ and extend the functional $\pi(M,\cdot)$ to $BV(\Theta)$ by defining
\begin{align*}
    \overline \pi(M,H) = \int_\Theta \left(v(\theta,x(\theta))-\int_{\underline \theta}^{\theta}v_1(s,x(s)) \diff s-c(x(\theta))\right) \diff H(\theta), \quad \forall H \in BV(\Theta).
\end{align*}
Clearly, $\overline \pi(M, F) = \pi(M, F)$ for all $F \in \mathcal F$. 
Moreover, note that for all $F, G \in BV(\Theta)$, 
\begin{align*}
|\bar\pi(M,F)-\bar\pi(M,G)|&=\left|\int_\Theta \left(v(\theta,x(\theta))-\int_{\underline \theta}^{\theta}v_1(s,x(s))ds-c(x(\theta))\right)\diff(F-G)(\theta)\right|\\
&\leq \left|\int_\Theta h_1\diff(F-G)(\theta)\right|+\left|\int_\Theta h_2(\theta)\diff(F-G)(\theta)\right|
\end{align*}
where $h_1(\theta):=v(\theta,x(\theta))$ and $h_2(\theta):=\int_{\underline \theta}^{\theta}v_1(s,x(s))\diff s+c(x(\theta))$. 
As both $v$ and $x$ are nondecreasing, we have that, for any $x(\cdot)$, 
\begin{align*}
    V(h_1)=v(\overline \theta, x(\overline \theta))-v(\underline \theta, x(\underline \theta)) \leq v(\overline \theta, \overline x)=:L_1<\infty.
\end{align*}
As $v$ is supermodular and nondecreasing in $\theta$ and $c$ is increasing and convex, 
\begin{align*}
    V(h_2)&=\int_{\underline \theta}^{\overline \theta}v_1(s,x(s)) \diff s+c(x(\overline \theta))-c(x(\underline \theta))
    \leq (\overline \theta - \underline \theta) \cdot \max_{\theta' \in \Theta}v_1(\theta', \overline x)+c(\overline x)=:L_2<\infty
\end{align*}
Moreover, we note that $\inf_{\theta \in \Theta} h_1(\theta)=v(\underline \theta,x(\underline \theta))=0$ and $\inf_{\theta \in \Theta} h_2(\theta)=\int_{\underline \theta}^{\underline \theta}v_1(s,x(s))\diff s+c(x(\underline \theta))= c(0)=0$. 
Hence, for $h_i$, with $m_i := \inf_{\theta \in \Theta} h_i(\theta)$, $i=1,2$, and letting $H(\theta)= F(\theta) - G(\theta)$,
\begin{align*}
    &\left|\int_{\underline \theta}^{\overline \theta} h_i(\theta) \diff F(\theta) - \int_{\underline \theta}^{\overline \theta} h_i(\theta) \diff G(\theta)\right| = \left|\int_{\underline \theta}^{\overline \theta} h_i(\theta) \diff [F(\theta) - G(\theta)]\right| \\
    &\leq \max\left\{m_i [H(\overline \theta) - H(\underline \theta)] + V(h_i) \sup_{\underline \theta \leq \alpha < \beta \leq \overline \theta} \int_\alpha^\beta \diff H\,\,,\,\, -m_i [H(\overline \theta) - H(\underline \theta)] - V(h_i) \inf_{\underline \theta \leq \alpha < \beta \leq \overline \theta} \int_\alpha^\beta \diff H\right\}\\
    &= V(h_i)\cdot \max\left\{\sup_{\underline \theta \leq \alpha < \beta \leq \overline \theta} \int_\alpha^\beta \diff H \,\,,\,\,\ \sup_{\underline \theta \leq \alpha < \beta \leq \overline \theta} -\int_\alpha^\beta \diff H\right\} \\
    &= V(h_i)\cdot \max\left\{\sup_{\underline \theta \leq \alpha < \beta \leq \overline \theta} H(\beta) - H(\alpha)\,\,,\,\,\ \sup_{\underline \theta \leq \alpha < \beta \leq \overline \theta} H(\alpha) - H(\beta)\right\} \\
    &\leq  V(h_i) \cdot \sup_{\underline \theta \leq \alpha < \beta \leq \overline \theta} |H(\alpha)| + |H(\beta)| = 2 \cdot V(h_i) \cdot {\|F - G\|}_\infty
\end{align*}

Combining the preceding inequalities results in
$\displaystyle|\bar\pi(M,F)-\bar\pi(M,G)| \leq 2(L_1 + L_2){\|F - G\|}_\infty$, $\forall F, G \in BV(\Theta), \; M \in \mathcal M$. 
The result is then obtained by restricting the domain to $\mathcal F$.

\subsection{Proof of \hyref{theorem:optimal-profit:frechet}[Theorem]}
\label{appendix:theorem:optimal-profit:frechet}

Let $\mathcal Y$ be a normed vector space with an open subset $\mathcal A \subseteq \mathcal Y$. 
As is standard, we denote the dual space of $\mathcal Y$ by $\mathcal Y^*$.
Before we prove the main result, let us recall two different notions of differentiability:

\begin{definition}[Gateaux Differentiability]
    A functional $T: \mathcal Y \to \mathbb R$ is Gateaux differentiable at $F \in \mathcal Y$ if there exists a linear functional $\dot T_F \in \mathcal Y^*$ such that for all $H \in \mathcal Y$,
    \begin{align*}
        \dot T_F(H) = \lim_{t \to \infty}\frac{T(F + tH)-T(F)}{t}.
    \end{align*}
\end{definition}

\begin{definition}[Fr\'{e}chet Differentiability]
    A functional $T: \mathcal Y \to \mathbb R$ is Fr\'{e}chet differentiable at $F \in \mathcal Y$ if there is a linear continuous functional $\dot T_F$ defined on $\mathcal Y$ such that 
    \begin{align*}
        \lim_{\|H\|\to 0}\frac{|T(F + H)-T(F)-\dot T_F(H)|}{\|H\|} = 0.
    \end{align*}
\end{definition}
Note that if $T$ is Gateaux differentiable at $F \in \mathcal Y$, then its derivative is unique. 
Moreover, if it is Fr\'{e}chet differentiable at $F \in \mathcal Y$, it is Gateaux differentiable and its derivatives agree. 

An important generalization of Gateaux differential for the case of convex and continuous functionals is that of a subdifferential.
\begin{definition}
    The subdifferential of $T: \mathcal Y \to \mathbb R$ at $F \in \mathcal Y$ is the set
    \begin{align*}
        \partial T(F) := \left\{D \in \mathcal Y^*: T(G)\geq T(F)+D(G-F)\text{ for each }G \in \mathcal Y\right\}.
    \end{align*}
\end{definition}

Since the normed linear space $(BV(\Theta),\|\cdot\|_\infty)$ is a metric space, and thus $BV(\Theta)$ is open, the following lemma guarantees that $\partial T$ is nonempty:
\begin{lemma}[Theorem 3.3.1. \citet{NiculescuPersson2018}]
    \label{lemma:subdifferential}
    If $T: \mathcal A \to \mathbb R$ is a continuous and convex functional, then $\partial T(a) \ne \emptyset$, for all $a \in \mathcal A$.
\end{lemma}

The next result shows why subdifferentials can be thought of as a generalization of Gateaux differentials. 
\begin{lemma}[Proposition 3.6.9. \citet{NiculescuPersson2018}]
    \label{lemma:subdifferential-Gauteaux}
    Let $T: \mathcal A \to \mathbb R$ be a continuous and convex functional.
    $T$ is Gateaux differentiable at $a \in \mathcal A$ if and only if $\partial T(a)$ is a singleton.
\end{lemma}

In what follows, we borrow from the proof strategy of Theorem 2 in \citet{BattauzDonnoOrtu2015MathFinEcon}, although their conditions do not directly apply to our problem.

First fix $M \in \mathcal M$ and notice that $\overline \pi(M,\cdot): BV(\Theta) \to \mathbb R$ is a linear functional. 
Therefore, it has a Fr\'{e}chet derivative. In fact, for all $F,H \in BV(\Theta)$,
\begin{align*}
    &\lim_{\|H\|_{\infty} \to 0} \frac{|\bar\pi(M, F + H) - \bar\pi(M,F) - \bar\pi(M, H)|}{\|H\|_{\infty}}
    \quad= \lim_{\|H\|_{\infty} \to 0} \frac{|\bar\pi(M, F) + \bar\pi(M,H) - \bar\pi(M,F) - \bar\pi(M, H)|}{\|H\|_{\infty}} = 0.
\end{align*}
This implies that the Fr\'{e}chet derivative of $\bar\pi(M,F)$ is independent of $F$ and given by $\dot \pi_M(\cdot) = \bar\pi(M,\cdot)$.
Therefore, $\bar\pi(M,\cdot)$ is also Gateaux differentiable, with Gateaux derivative at $F$ given by $\bar\pi(M,\cdot)$.

Define the functional $\overline \Pi: BV(\Theta) \to \mathbb R$ by
$\displaystyle\overline \Pi(H) = \sup_{M \in \mathcal M} \overline \pi(M,H)$, $\forall H \in BV(\Theta)$.
Since the set $\left\{\overline \pi(M,H) : M \in \mathcal M\right\}$ is bounded for any $H \in BV(\Theta)$, it is clear that $\overline \Pi(H)<\infty$.
Furthermore, it is immediate that $\overline \Pi(F) = \Pi(F)$ for all $F \in \mathcal F$.

As $\overline \pi(M,H)$ is linear in $H$ for any $M$, one immediately has that $\bar\Pi$ is convex in $H \in BV(\Theta)$, as the supremum of a family of linear functionals.
Moreover, from the proof of \hyref{lemma:optimal-profit:lipschitz}[Lemma], it is easy to see that $\overline \Pi$ remains Lipschitz continuous. Therefore, by \hyref{lemma:subdifferential}[Lemma], $\partial \bar\Pi(H) \ne \emptyset$ for any $H \in BV(\Theta)$.

By \hyref{lemma:existence}[Lemma], $\mathcal M^*(F)\ne \emptyset$ for any $F \in \mathcal F$.
Fix any $F \in \mathcal F$. Take any $D \in \partial \bar\Pi(F)$ and any $M_F \in \mathcal M^*(F)$.
Note, for any $G \in BV(\Theta)$, that $\bar\pi(M_F,F)-\bar\pi(M_F,G)\geq \bar\Pi(F)-\bar\Pi(G)\geq D(F-G)$, hence $\partial \bar\Pi(F) \subseteq \partial \bar\pi(M_F,F)$ and, as $\bar\pi(M_F,\cdot)$ is continuous and linear, by \hyref{lemma:subdifferential-Gauteaux}[Lemma], $\partial \bar\pi(M_F,F) = \left\{\dot \pi_{M_F}\right\}$.
Then, for any $G \in BV(\Theta)$,
\begin{align*}
    &\bar\pi(M_F,F)-\bar\pi(M_F,G)\geq \bar\Pi(F)-\bar\Pi(G)\geq \dot \pi_{M_F,F}(F-G)\\
    \Longleftrightarrow \quad&\bar\pi(M_F,F)-\bar\pi(M_F,G)-\dot \pi_{M_F}(F-G)\geq \bar\Pi(F) - \bar\Pi(G)-\dot \pi_{M_F}(F-G)\geq 0\\
    \Longrightarrow \quad&|\bar\pi(M_F,F)-\bar\pi(M_F,G)-\dot \pi_{M_F}(F-G)|\geq |\bar\Pi(F)-\bar\Pi(G)-\dot \pi_{M_F}(F-G)|\geq 0.
\end{align*}
By Fr\'{e}chet differentiability of $\bar\pi(M_F,\cdot)$, we then have that $\forall {\{G_n\}}_n$ such that $\|G_n-F\|_\infty\to 0 $, 
\begin{align*}
    0&\leq \frac{|\bar\Pi(G_n)-\bar\Pi(F)-\dot \pi_{M_F}(G_n-F)|}{\|G_n-F\|_\infty}\leq \frac{|\bar\pi(M_F,G_n)-\bar\pi(M_F,F)-\dot \pi_{M_F}(G_n-F)|}{\|G_n-F\|_\infty} \to 0
\end{align*}
and, consequently, $\bar\Pi$ is Fr\'{e}chet differentiable at $F \in \mathcal F$.
As $F$ was arbitrary, we have that $\bar\Pi$ is Fr\'{e}chet differentiable at any $F \in \mathcal F$.

\subsection{Proof of \hyref{theorem:profit:asymptotic-normality}[Theorems] and \ref{theorem:optimal-profit:asymptotic-normality}}
\label{appendix:theorem:profit:asymptotic-normality}
\label{appendix:theorem:optimal-profit:asymptotic-normality}

We will prove \hyref{theorem:optimal-profit:asymptotic-normality}[Theorem]. The proof for \hyref{theorem:profit:asymptotic-normality}[Theorem] is virtually the same.

By \hyref{theorem:optimal-profit:frechet}[Theorem], $\Pi(F)$ is Fr\'{e}chet differentiable at any $F \in \mathcal F$ and, thus, it can be written as $\Pi(F)=\Pi(F_0)+\dot \Pi_{F_0}(F-F_0)+o(\|F-F_0\|_\infty)$.
Furthermore, we note that $\hat F$ is an unbiased estimator of $F_0$ and then, by linearity, 
\begin{align*}
    \mathbb E\left[\frac{1}{n}\sum_{i=1}^n\pi(M^*(F_0),\delta_{\theta_i})\right]&=
    \mathbb E\left[\pi(M^*(F_0),\hat F(S^n))\right]=\pi\left(M^*(F_0),\mathbb E\left[\hat F(S^n)\right]\right)=\pi(M^*(F_0),F_0),
\end{align*} 
where $\delta_{\theta}$ denotes the cumulative distribution function associated with a Dirac delta measure at $\theta$ and $\theta_i$ is the $i$-th observation in the sample $S^n$.
Thus,
\begin{align*}
    &\sqrt{n}\left(\Pi(\hat F(S^n)) - \Pi(F_0)\right)=
    \frac{1}{\sqrt n}\sum_{i=1}^n\left(\pi(M^*(F_0),\delta_{\theta_i})-\pi(M^*(F_0),F_0)\right) +\sqrt n \cdot o(\|\hat F(S^n)-F_0\|_\infty)\\
    &=\frac{1}{\sqrt n}\sum_{i=1}^n\left(\pi(M^*(F_0),\delta_{\theta_i})-\pi(M^*(F_0),F_0)\right)
     +\sqrt n\cdot \|\hat F(S^n)-F_0\|_\infty \cdot \frac{o(\|\hat F(S^n)-F_0\|_\infty)}{\|\hat F(S^n)-F_0\|_\infty}\\
    &=\frac{1}{\sqrt n}\sum_{i=1}^n\left(\pi(M^*(F_0),\delta_{\theta_i})-\pi(M^*(F_0),F_0)\right)+O_p(1) \cdot o(1) \quad \overset{d}{\to} N(0,\sigma_{F_0}^2),
\end{align*} 
where $\sigma_{F_0}^2=\mathbb E\left[{\left(\dot\Pi_{F_0}(\delta_{\theta}-F_0)\right)}^2\right]$. 

Finally, following \citet{Parr1985StatProbL}, we have that
\begin{align*}
    \hat G_n=&\sqrt{n}\left(\Pi(\hat F(S^n_B)) - \Pi(\hat F(S^n))\right)=\sqrt{n}\left(\left\{\Pi(F_0)+\dot\Pi_{F_0}(\hat F(S^n_B)-F_0)+o(\|\hat F(S^n_B)-F_0\|_\infty)\right\}\right. \\ &\hspace*{14em}\left. - \left\{\Pi(F_0)+\dot\Pi_{F_0}(\hat F(S^n)-F_0)+o(\|\hat F(S^n)-F_0\|_\infty)\right\}\right)\\
    =&\sqrt{n}\left(\dot\Pi_{F_0}(\hat F(S^n_B)-\hat F(S^n))+o(\|\hat F(S^n_B)-F_0\|_\infty)+o(\|\hat F(S^n)-F_0\|_\infty)\right)\\
    =&\sqrt{n}\left(\dot\Pi_{F_0}(\hat F(S^n_B)-\hat F(S^n))+o(\|\hat F(S^n_B)-\hat F(S^n)\|_\infty)+o(\|\hat F(S^n)-F_0\|_\infty)\right),
\end{align*}
where we used Fr\'{e}chet differentiability to obtain a first-order von Mises expansion of $\Pi$ in the second equality, linearity of $\dot\Pi_{F_0}$ in the third and the triangle inequality in the last.
As, $\hat F(S^n_B)$ and $\hat F(S^n)$ denote empirical distributions of $\hat F(S^n)$ and $F_0$, by the \citepos{DvoretzkyKieferWolfowitz1956AnnMathStat} inequality, we have that
\begin{align*}
    \hat G_n&=\sqrt{n}\dot\Pi_{F_0}(\hat F(S^n_B)-\hat F(S^n))+o_p(1)=\frac{1}{\sqrt n}\sum_{i=1}^n\left(\pi(M^*(F_0),\delta_{\theta_i^B})-\pi(M^*(F_0),\hat F(S^n))\right)+o_p(1)
\end{align*}
As $\mathbb E\left[\pi(M^*(F_0),\delta_{\theta_i^B})\mid S^n\right]=\pi(M^*(F_0),\hat F(S^n))$ and $\mathbb E\left[{\left(\pi(M,\hat F(S^n_B))\right)}^2\right]=\sigma_{F_0}^2<\infty$, by the central limit theorem in \citet{BickelFreedman1981AnnStat}, we have that $\hat G_n \overset{d}{\to} G_0$.
Finally, as the limiting distribution is continuous, convergence is uniform.

\subsection{Proof of \hyref{proposition:optimal-profit:liecdf}[Proposition]}
\label{appendix:proposition:optimal-profit:liecdf}

Let $\hat F^E$ denote the empirical distribution estimator.
By \hyref{lemma:optimal-profit:lipschitz}[Lemmas] and \ref{lemma:iecdf}, $\Pi(\hat F(S^n))=\Pi(\hat F^E(S^n))+O_p(n^{-1})$.
As such, (1) follows from the observation that $\sqrt n\left\{\Pi(\hat F(S^n)) - \Pi(F_0)\right\} \\= \sqrt n\left\{\Pi(\hat F^E(S^n))-\Pi(F_0)\right\} + O_p(n^{-1/2})$, which, together with Slutsky's theorem and \hyref{theorem:optimal-profit:asymptotic-normality}[Theorem] implies $\sqrt n\left\{\Pi(\hat F(S^n))-\Pi(F_0)\right\}\overset{d}{\to}N(0,\sigma^2_{F_0})$. 
(2) results from the analogous observation that:
\begin{align*}
    \hat G_n&=\sqrt{n}\left(\Pi(\hat F(S^n_B)) - \Pi(\hat F(S^n))\right)=\sqrt{n}\left(\Pi(\hat F^E(S^n_B)) - \Pi(\hat F^E(S^n))\right)+O_p(n^{-1/2})\overset{d}{\to} N(0,\sigma_{F_0}^2).
\end{align*}
For (3), we first prove the following lemmas:
\begin{lemma}
    \label{theorem:mconv}
    Consider distribution functions $F, \; F_1, \; F_2, \ldots$ such that $\|F_n(t) - F(t)\|_\infty \xrightarrow{a.s.} 0$ for all $t \in \mathbb R$. 
    For all $n \in \mathbb N$, let $\mu_n$ be the measure on $(\mathbb R,\mathcal B(\mathbb R))$ induced by $F_n$ and $\mu$ be the measure induced by $F$. 
    Then $\mu_n(A) \to \mu(A)$ for all $A \in \mathcal B(\mathbb R)$ almost surely.
\end{lemma}
\begin{proof}
    Let $\left\{I_j := (a_j, b_j] \subseteq \mathbb R \;|\; j \in \mathbb N\right\}$ be a collection of disjoint intervals, and denote $I = \bigcup_{j = 1}^\infty I_j$. 
    Then, since for any $n \in \mathbb N$, $\mu_n$ and $\mu$ are finite measures,
    \begin{align*}
        & \left|\mu_n(I) - \mu(I)\right| = \left|\sum_{j = 1}^\infty \mu_n(I_j) - \sum_{j = 1}^\infty \mu(I_j)\right|= \left|\sum_{j = 1}^\infty F_n(b_j) - F_n(a_j) - \sum_{j = 1}^\infty F(b_j) - F(a_j)\right| \\
        &= \left|\sum_{j = 1}^\infty \left(F_n(b_j) - F(b_j)\right) - \left(F_n(a_j) - F(a_j)\right)\right| \leq \sum_{j = 1}^\infty\left|F_n(b_j) - F(b_j)\right| + \left|F_n(a_j) - F(a_j)\right| \leq \sum_{j = 1}^\infty 2 \sup_{t \in \mathbb R} \left|F_n(t) - F(t)\right|
    \end{align*}
    
    Let $c_n = \sup_{t\in\mathbb R}|F_n(t) - F| \to 0$. 
    Then, there exists a monotone convergent subsequence $c_{n_k} = 2\sup_{t \in \mathbb R}\left|F_{n_k}(t) - F(t)\right| \to 0$ as $k \to \infty$. 
    Therefore, by the monotone convergence theorem, 
    \[\lim_{k\to \infty}\left|\mu_{n_k}(I) - \mu(I)\right| \leq \lim_{k \to \infty} \sum_{j=1} 2\sup_{t\in \mathbb R}\left|F_{n_k}(t) - F(t)\right| = \sum_{j=1} \lim_{k \to \infty} 2\sup_{t\in \mathbb R}\left|F_{n_k}(t) - F(t)\right| = 0.\]
    This implies there exists a convergent subsequence $\mu_{n_k}(I) \to \mu(I)$.

    Now take any convergent subsequence $\mu_{n_m}(I)$ of $\mu_n(I)$, which we will denote by $\mu_m(I)$. 
    We have
    \[ \left|\mu_m(I) - \mu(I)\right| \leq \sum_{j=1}^\infty 2\sup_{t \in \mathbb R}\left|F_m(t) - F(t)\right|.\]
    Then $c_m = 2\sup_{t \in \mathbb R}\left|F_m(t) - F(t)\right| \to 0$ has a monotone convergent subsubsequence $c_{m_r}$. 
    We can thus apply the monotone convergence theorem once again to conclude that
    $\displaystyle\left|\mu_{m_r}(I) - \mu(I)\right| \to 0$ as $r \to \infty$. 
    Since the subsequence $\mu_m(I)$ is convergent by assumption, it must converge to the limit of each of its subsequences, and we have
    $\displaystyle\lim_{m \to \infty}\mu_{m}(I) = \mu(I)$.

    Therefore, every convergent subsequence of $\mu_n(I)$ converges to $\mu(I)$. 
    Since ${\left(\mu_n(I)\right)}_{n \in \mathbb N}$ is bounded, this implies that $\lim_{n \to \infty} \mu_n(I) = \mu(I)$, and thus $\lim_{n \to \infty} \mu_n$ is a pre-measure that agrees with $\mu$ in the ring formed by disjoint unions of intervals of the type $(a,b], \; b > a$. 
    Therefore, since $\mu$ is $\sigma$-finite, Carath\'{e}odory's extension theorem implies that $\lim_{n \to \infty} \mu_n$ must agree with $\mu$ on $\mathcal B(\mathbb R)$ almost surely.
\end{proof}

\begin{lemma}
    Let $\left\{F_n: \Theta \to [0,1] \;|\; n \in \mathbb N \right\}$ be a sequence of absolutely continuous distribution functions with Radon-Nikodym derivatives given by $f_n$. 
    If there exists a distribution function $F$ such that ${\|F_n - F\|}_\infty \to 0$, then it is absolutely continuous with Radon-Nikodym derivative $f$ and $\|f_n-f\|_1\to 0$.
\end{lemma}
\begin{proof}
    As $\forall n \in \mathbb N$, $F_n$ is absolutely continuous and $\Theta$ has finite Lebesgue measure, then the Radon-Nikodym derivatives $\{f_n\}_{n\in \mathbb N}$ are uniformly integrable with respect to the Lebesgue measure.
    Let $\mu$ be the measure associated with $F$. 
    By the Vitali-Hahn-Saks theorem and \hyref{theorem:mconv}[Lemma], we have that
    \begin{align*}
        \lim_{n \to \infty} \int_A f_n(\theta) \diff \theta = \mu(A)
    \end{align*}
    for all $A \in \mathcal B(\Theta)$. 
    Since $\{f_n\}_{n\in \mathbb N}$ is uniformly integrable, the Dunford-Pettis theorem implies that every subsequence of $\{f_n\}_{n\in \mathbb N}$ has a convergence subsubsequence converging to $g$ in $L^1(\Theta)$. 
    Denote such a subsubsequence by ${\{f_{n_k}\}}_{k \in \mathbb N}$. 
    Then, for every $A \in \mathcal B(\mathbb R)$,
    \begin{align*}
        \int_A g(\theta)\diff \theta = \lim_{k \to \infty} \int_A f_{n_k}\diff \theta = \lim_{k \to \infty} \mu_{n_k}(A) = \int_A f(\theta)\diff \theta.
    \end{align*}
    This implies that $f = g$ almost surely and ${\|f_n - f\|}_1 \to 0$.
\end{proof}
Given that, ${\{\hat F(S^n)\}}_{n\in \mathbb N}$ is absolutely continuous with probability 1 and ${\|\hat F(S^n)-F_0\|}_\infty \overset{a.s.}{\to}0$, the previous lemmas imply that ${\|f_n-f_0\|}_1\overset{p}{\to} 0$, which concludes the proof for (3).

\subsection{Proof of \hyref{proposition:auctions}[Proposition]}
\label{appendix:proposition:auctions}

First, we note that $\pi(r,F)$ is Lipschitz continuous in $F$, with a Lipschitz constant that is independent of $r$.
Note that
\begin{align*}
    & |\pi(r,F)-\pi(r,G)|=\left|\int_\Theta \mathbf 1_{\theta \geq r}\diff(F_{(2;M)}-G_{(2;M)})\right| \leq \|F_{(2;M)}-G_{(2;M)}\|_\infty\\
    &=\sup_{\theta \in \Theta}|M\cdot (F(\theta)^{M-1}-G(\theta)^{M-1})+(M-1)\cdot (G(\theta)^M-F(\theta)^{M})|\\
    &\leq M\cdot \sup_{\theta \in \Theta}|F(\theta)^{M-1}-G(\theta)^{M-1}|+(M-1)\cdot \sup_{\theta \in \Theta}|F(\theta)^M-G(\theta)^{M})|\\
    &\leq 2M(M-1)\|F-G\|_\infty.
\end{align*}
where the first inequality uses the Beesack-Darst-Pollard inequality --- see \hyref{lemma:boundsintegral}[Lemma].
By the same arguments as in \hyref{lemma:optimal-profit:lipschitz}[Lemma], we have that $\Pi(F):=\sup_{r \in \Theta}\pi(r,F)$ is also Lipschitz continuous in $F$ and, by those made in \hyref{proposition:asymptotic-optimality}[Propositions] and \ref{proposition:regretbounds}, the result follows.

\subsection{Other Proofs}
\label{appendix:other-proofs}

Let the linearly interpolated empirical cumulative distribution be given by
\begin{align*}
    \hat F(S^n)(\theta)=\sum_{k=0}^{n-1}\mathbf 1_{\{\theta_{(k)}\leq \theta<\theta_{(k+1)}\}}\frac{1}{n}\frac{\theta-\theta_{(k)}}{\theta_{(k+1)}-\theta_{(k)}}+\mathbf 1_{\{\theta_{(n)}\leq \theta\}},
\end{align*}
where $\theta_{(k)}$ denotes the $k$-th smallest observation in the sample $S^n$ and $\theta_{(0)}=\underline \theta$.
The following holds:
\begin{lemma}
    \label{lemma:iecdf}
    For any absolutely continuous $F_0 \in \mathcal F$, (1) $\|\hat F(S^n)-F_0\|_\infty \overset{a.s.}{\to}0$ and (2) with probability 1, $\hat F(S^n) \in \mathcal F$, $\hat F(S^n)$ has convex support and is absolutely continuous.
\end{lemma}
\begin{proof}
    Note that, with probability 1, any sampled value $\theta_i>\underline \theta$ as $F_0$ is absolutely continuous and so $\hat F(S^n)$ is well defined.
    By construction, $\hat F(S^n)$ has convex support and is absolutely continuous.
    As the probability that any two sampled observations have the same value is null, we have that, with probability 1, ${\|\hat F(S^n)-\hat F^E(S^n)\|}_\infty = 1/n$ where $\hat F^E$ denotes the empirical cumulative distribution and, with probability 1, ${\|\hat F(S^n)-F_0\|}_\infty= {\|\hat F^E(S^n)-F_0\|}_\infty+1/n$.
    Consequently, as ${\|\hat F^E(S^n)-F_0\|}_\infty \overset{a.s.}{\to}0$ and as $F_0$ is absolutely continuous, then the result follows.
\end{proof}

\end{document}